\documentclass[3p,times,review,numbers,authoryear]{elsarticle}

\usepackage{graphicx}
\usepackage{caption}
\usepackage{subfig}
\usepackage{array}
\usepackage{mathtools}
\usepackage[mathscr]{euscript}
\usepackage[shortlabels]{enumitem}
\usepackage{float}
\usepackage{epstopdf}
\usepackage{algorithm}
\usepackage{stmaryrd}
\usepackage{algpseudocode}
\usepackage{multirow}
\usepackage{booktabs}
\usepackage{url}
\usepackage{hhline}
\usepackage{color} 
\usepackage{amssymb}
\usepackage[numbers]{natbib}
\usepackage[final]{pdfpages}

\newtheorem{thm}{Theorem}

\newdefinition{rmk}{Remark}
\newproof{pf}{Proof}

\begin{document}
	
\begin{frontmatter}

This is a preprint of the paper:

Amna Qureshi, David Meg\'{i}as, Helena Rif\`a{}-Pous, ``PSUM: Peer-to-peer multimedia content distribution
using collusion-resistant fingerprinting", Journal of Network and Computer Applications, Volume 66, $31^{\mathit{st}}$ May 2016, Pages 180-197, ISSN 1084-8045, \url{http://dx.doi.org/10.1016/j.jnca.2016.03.007}.

\newpage

\title{PSUM: Peer-to-peer multimedia content distribution using collusion-resistant fingerprinting}


\author{Amna Qureshi, David Meg\'{i}as, Helena Rif\`a{}-Pous}
\address{Estudis d'Inform\`{a}tica Multim\`{e}dia i Telecomunicaci\'{o},\linebreak
	Internet Interdisciplinary Institute (IN3), \linebreak
	Universitat Oberta de Catalunya (UOC),\linebreak
	Barcelona, Spain.\linebreak
	Email:\{aqureshi,dmegias,hrifa\}@uoc.edu}

\begin{abstract}
	The use of peer-to-peer (P2P) networks for multimedia distribution has spread out globally in recent years. The mass popularity is primarily driven by cost-effective distribution of content, also giving rise to piracy. An end user (buyer/peer) of a P2P content distribution system does not want to reveal his/her identity during a transaction with a content owner (merchant), whereas the merchant does not want the buyer to further distribute the content illegally. To date, different P2P distribution systems have been proposed that provide copyright and privacy protection at a cost of high computational burden at the merchants and/or at the buyer’s end and thus, making these systems impractical. In this paper, we propose PSUM, a P2P content distribution system which allows efficient distribution of large-sized multimedia content while preserving the security and privacy of merchants and buyers. The security of PSUM is ensured by using an asymmetric fingerprinting protocol based on collusion-resistant codes. In addition, PSUM enables buyers to obtain digital contents anonymously, but this anonymity can be revoked as soon as he/she is found guilty of copyright violation. The paper presents a thorough performance analysis of PSUM, through different experiments and simulations, and also analyzes several security compromising attacks and countermeasures.
\end{abstract}


\end{frontmatter}

\section{Introduction}
\label{sec:introduction}
Traditional client-server content distribution systems are dependent on a centralized server which is costly in terms of initial infrastructure investment and maintenance. Moreover, the lack of scalability and the high bandwidth requirements are some factors that degrade the client-server system performance. In contrast to client-server systems, P2P technology offers cost efficiency, scalability, less administrative requirements and exposure to a large number of users. These benefits are the attractive features for media companies towards the adoption of P2P systems. BitTorrent (BT) \citep{BitTorrent} is one of the most popular P2P distribution systems used on the Internet for distributing large amount of data, and it accounts for a significant volume of Internet traffic. For example, Red Hat Inc. uses BT to distribute Red Hat Linux. Also, many open source software, game and new media companies use BT for the distribution of software, game updates and videos, respectively.

Despite P2P content distribution technology has the potential to revolutionize the Internet in numerous respects, it has often been surrounded with the copyright controversy. The copyright holders encounter uncertainties regarding the adoption or rejection of P2P networks to spread content over the Internet. They apparently fear losing control of content ownership and worry about the illegal activity promotion. Also, the decentralized nature of the P2P technology makes tracing \citep{cfn94} of a copyright violator a challenging task. Therefore, mechanisms must be deployed to ensure that the multimedia content can be used safely by legitimate end users. Encryption can provide multimedia data with desired security during transmission by preventing them from unauthorized access. However, it cannot prevent an end user from re-distributing the data illegally once it has been received and decrypted. As one of the promising solutions, digital fingerprinting addresses the problems of copyright protection, tamper detection and traitor tracing.

In digital fingerprinting, a buyer-specific identification mark, known as a fingerprint, is embedded into different copies of the same content. The resulting copies are referred to as fingerprinted copies and each fingerprinted copy is assigned to a buyer. In traditional fingerprinting, known as symmetric fingerprinting, the fingerprint is generated and embedded solely by the merchant and the buyer has no control over the embedding process \citep{ckls97}. Thus, a dishonest merchant could frame an innocent buyer (customer’s rights problem), while a cheating buyer would be able to deny his/her responsibility for a copyright violation act (non-repudiation problem). Asymmetric fingerprinting schemes \citep{ps96} were introduced to overcome this shortcoming. In that case, a buyer chooses a secret and sends a commitment to the secret to a merchant. Then through an interactive protocol with the merchant, the buyer obtains a fingerprinted content with his/her secret, while the merchant does not have access to the fingerprinted content obtained by the buyer. In case the merchant finds a pirated copy, he/she can extract the secret chosen by the buyer, identify him/her, and prove he/she is guilty of re-distribution. However, most of the current asymmetric fingerprinting schemes in the literature incur high computational and communicational burdens at the merchant’s and/or at the buyer’s end, due to the use of cryptographic protocols such as homomorphic encryption or committed oblivious transfer.

\textbf{\textit{Our contribution}}: The main contribution of this paper is to introduce an asymmetric fingerprinting protocol with a novel design that provides a secure, anonymous and efficient collusion-resistant-based fingerprinting scheme within a P2P content distribution system. We have proposed a system, Privacy and Security of User and Merchant (PSUM), that aims to provide the following properties:

\begin{enumerate}
	\item To perform efficiently in a decentralized network, differing from the existing fingerprinting schemes, the heavy-burden operations, such as public-key encryption, are restricted only to key exchanging, data signing and encryption of small-length strings. Furthermore, the proposed scheme reduces the computational and communicational costs of the merchant by using the idea of file partitioning. The multimedia file is partitioned by the merchant into a small-sized base file and a large-sized supplementary file. The base file contains the most important information and, without it, the supplementary file is unusable. The merchant sends the base file to a buyer in a semi-centralized way and uses a network of peer buyers to distribute the supplementary file.
	\item The proposed asymmetric fingerprinting protocol based on a state-of-the-art collusion-resistant codes and an existing secure embedding scheme is performed between a merchant, a buyer and a set of P2P proxies in the presence of a third party (monitor). The proposed fingerprinting protocol (Section \ref{sec:3.4.2}) provides significantly improved efficiency over that of similar schemes that have been presented in the past, by using the idea of pre-computation-based secure embedding and permutation. In pre-computation-based secure embedding, a merchant applies the Discrete Wavelet Transform (DWT) to a multimedia content to obtain low-frequency (approximation) coefficients to form a base file. These approximation coefficients are embedded in parallel with all 1s and all 0s bit streams using a state-of-the art robust and secure watermark embedding scheme. The two variants of the approximation coefficients form a base file, which is then distributed to a buyer through a protocol providing asymmetric fingerprinting. The permutation construct used in our fingerprinting protocol provides non-repudiation and protection against the customer’s right problem unlike existing fingerprinting protocols, which require highly demanding technologies such as homomorphic cryptosystems, secure multi-party computation protocols, zero-knowledge proofs, among others. The permutation construct enables the merchant to reduce the bandwidth and CPU time unlike existing fingerprinting protocols, which employ homomorphic cryptosystems and secure multi-party computation protocols that results in expansion of the content and increased computational costs. The fingerprint generated by the trusted monitor is permuted using different keys and is then assigned to a set of proxy peers in such a way that the merchant cannot predict about the fingerprint and the fingerprinted content, and the proxy peers are unable to frame honest buyers by combining their information bits.
	\item The state-of-the art collusion-resistant fingerprinting codes used in our proposed protocol can prevent the collusion of a set of malicious buyers who intend to remove or alter the fingerprint from the content so as to evade being traced and at the same time possibly frame an innocent buyer. Also, it provides traceability, which im- plies that the extracted pirated codeword from the pirated copy can be used to identify the buyer who is found responsible for illegal re-distribution of the content.
	\item The proposed scheme can protect the buyer’s anonymity until he/she is found guilty of illegal re-distribution as detailed in traitor-tracing protocol (Section \ref{sec:3.4.4}). Anonymity is, thus conditional and revocable in PSUM. The anonymity to the buyer is provided by using dynamic pseudonyms based on a one-way hash function and the revocability is achieved by successful run of the adopted collusion-resistant fingerprinting scheme’s \citep{Nu10} tracing protocol in conjunction with the proposed arbitration and identification protocol. The tracing protocol of \cite{Nu10} codes outputs the pseudonym of an allegedly guilty buyer, which is then adjudicated by the judge either guilty or innocent. If the buyer is found guilty, his/her real identity is revealed by the certification authority of PSUM.
\end{enumerate}

\textbf{\textit{Outline of the paper}}: The remainder of this paper is organized as follows. Section \ref{sec:previous} reviews the related work on digital fingerprinting, collusion-resistant fingerprinting and P2P content distribution systems. The proposed P2P content distribution system, PSUM, is detailed in Section \ref{sec:PSUM}. In Section \ref{sec4}, the security analysis of PSUM is discussed in terms of privacy and security. The results of the experiments designed to evaluate the performance of PSUM are presented in Section \ref{sec5}. Finally, Section \ref{sec6} summarizes the conclusions.

\section{Related work}
\label{sec:previous}

This section reviews related work on asymmetric fingerprinting, collusion-resistant fingerprinting and P2P content distribution systems.

\subsection{Asymmetric fingerprinting}
\label{sec:asym}

Various asymmetric fingerprinting schemes have been proposed \citep{MaSeDoSo03,k10} in which the requirement of fair multimedia content distribution has become prevalent. Some asymmetric fingerprinting protocols also provide buyers with anonymity \citep{pw97,ps99,mw01}, in which trusted third parties are usually introduced to
provide fairness and anonymity to the merchant and the buyer, respectively. Various fingerprinting schemes do not involve trusted parties for the execution of the protocols \citep{csp03,dp08}.

Initial asymmetric fingerprinting protocols were based on bit-commitment schemes \citep{ps96,bm02}, which require high enciphering rates to achieve security. Thus, the implementation of these protocols involves a large overhead and high communicational cost. Other proposals, like \citep{kt05}, apply a homomorphic property of public-key cryptosystems to achieve asymmetric fingerprinting. The homomorphic property allows the merchant to embed the fingerprint in the encrypted domain in such a way that only the buyer obtains the decrypted fingerprinted content. However, the use of homomorphic encryption expands data and substantially increases the communication bandwidth required for data transfers. \cite{hl10} proposed an asymmetric fingerprinting protocol from the communication point of view and is based on a 1-out-of-2 oblivious transfer protocol. Thus, in any case, all the proposed asymmetric fingerprinting schemes involve complex cryptographic protocols which require high bandwidth and heavy computational costs. This makes the schemes impractical in a real-world scenario. \cite{pg99} prove that efficient fair exchange protocols cannot be completely fair without the help of a third party that is mutually trusted by both of the parties performing the exchange. Thus, using a trusted third party is a price worth paying if it can turn asymmetric fingerprinting scheme into a practical alternative \citep{MaSeDoSo03}.

\subsection{Collusion-resistant fingerprinting}
\label{sec:coll}

In digital fingerprinting, a unique fingerprint is embedded in each distributed copy that is used to trace and identify the source of illicit copies. However, due to the uniqueness of each distributed copy, digital fingerprinting systems are vulnerable to collusion attacks. In these attacks, the colluders can combine the information from different copies and generate a new copy in which the original fingerprints are either removed or attenuated. Much work on collusion- secure ($c$-secure) fingerprinting has been proposed in the literature \citep{cfn94,bm02,tspl03}. A code $F$ is totally $c$-secure if there exists a $c$-frameproof code and a tracing algorithm. In $c$-frameproof codes, no collusion of at most $c$ buyers can frame a buyer who is not a member of the collusion. The tracing algorithm is used when the merchant finds a pirated copy and wants to trace the members of the collusion $w$. The most well-known $c$-secure code is the \cite{bs99} (B-S) scheme. The code length of the B-S scheme is $O(c^4 \mathrm{log}(M/\in) \mathrm{log}(1/\in))$, where $M$ is the number of buyers, $c$ is the number of colluders and $\in$ is the probability of accusing an innocent buyer. Unfortunately, the large length of the B-S code restricts the range of its practical applications. Much research has been carried out to reduce the code length and improve its performance \citep{sd02,SCMA05}. \cite{t03} proposed a probabilistic fingerprinting code with theoretically minimal length $m = 100c^2\mathrm{log}(1/\in)$ with respect to the number of colluders $c$. Afterwards, many improved works have appeared from different directions. \cite{nfhkwoi07} presented a discrete version of the Tardos codes in an attempt to reduce the code length and memory requirements. These codes are based on a relaxed marking assumption (called $\delta$-marking assumption), in which the number of undetectable bits that are either erased or flipped is bounded by a $\delta$-fraction of the total code length. Except for the bias distribution, the Nuida et al.’s codes employ the same mechanism as the Tardos codes and are shorter in length. \cite{Nu10} proposed a new construction of collusion-secure fingerprint codes against up to three colluders. The fingerprint generation algorithm is similar to Tardos codeword generation algorithm except for the bias distribution. The novelty is the traitor-tracing algorithm, which combines the use of score computation analogous to Tardos codes with an extension of parent search technique.

However, merging collusion-resistant fingerprinting schemes and secure embedding is a difficult task. Early secure watermark embedding schemes \citep{kt05,dp08} assumed that the use of anti-collusion codes make the schemes resistant against collusion attacks without giving any proof-of-concept. Recently, two asymmetric fingerprinting schemes based on $c$-secure codes were proposed. \cite{cff11} proposed a solution that allows a buyer to pick up fingerprint bits from a list controlled by the merchant, in such a way that he/she does not know the chosen elements. However, the proposed scheme requires heavy computation due to use of an oblivious transfer protocol. Also, the number of communication rounds between a buyer and a seller is impracticable as it has a linear relation with the length of the code. \cite{p13} proposed an asymmetric fingerprinting scheme based on the B-S code with constant communication round but at a cost of a longer codeword.

\subsection{P2P content distribution systems}
\label{sec:P2P}

Many overlay networks have been proposed in recent years amongst which P2P networks are the most commonly applied. However, today’s P2P content distribution systems are severely abused by illegal re-distributions. Many systems can be found in the literature that incorporate content protection mechanisms to solve the copyright infringement problem in P2P systems. However, a collection of identifiable personal data within P2P systems using copyright protection mechanisms raises a privacy concern among the end users. The literature review shows that very few researchers have worked on a P2P content distribution system that provides preservation of content providers’ owner- ship properties and content receivers’ privacy, so far. The P2P content distribution systems described in the following paragraph satisfy both copyright protection and end user privacy concerns.

\cite{md13} proposed a P2P content distribution system that utilizes the fingerprinting concept to provide identification to the copyright owner, offers collusion resistance against dishonest buyers and detects illegal content re-distributors. In the tracing process, some peers have to cooperate in tracing a copyright violator. The buyers can also preserve their privacy as long as they do not get involved in illegal re-distribution. However, this system is implemented with a two-layer anti-collusion code (segment level and fingerprint level), which
results in a longer codeword. Furthermore, honest and committed proxies are required in a system of \cite{md13} for the generation of valid fingerprints as compared to PSUM which only requires an honest monitor for the fingerprint generation. \cite{Me14} proposed an improved version of the system proposed by \cite{md13} in which malicious proxies are considered in the fingerprinting protocol. A four-party anonymous communication protocol is proposed to prevent malicious proxies to access clear-text fingerprinted content and avoids graph search for traitor tracing. However, the system of \cite{Me14} still requires a two-layer anti-collusion code. \cite{dm13} proposed a P2P protocol for distributed multicast of fingerprinted content in which cryptographic primitives and a robust watermarking technique are used to produce different marked copies of the content for the requesting buyer such that it can help the provider to trace re-distributors without affecting the privacy of honest buyers. However, an implementation of a secure multi-party protocol results in increased computational and communication costs at the buyer end. \cite{qmr15} proposed a P2P content distribution framework for preserving privacy and security of the user and the merchant based on homomorphic encryption. In that framework, some discrete wavelet transform (DWT) low-frequency (approximation) coefficients are selected according to a secret key for embedding an encrypted fingerprint to prevent data expansion due to homomorphic encryption. Although the selective public-key encryption of the multimedia content results in lesser data expansion compared to encrypting the whole content, it imposes computational burden on a merchant and an increased complexity in file reconstruction at the buyer’s end.

Unlike the P2P content distribution systems described in the above paragraph, the following P2P distribution systems fail to provide privacy to the end users. A fingerprint generation and embedding method was proposed by \cite{lkm10} for complex P2P file sharing networks for copyright protection. In this system, wavelet transforms and principal component analysis (PCA) techniques are used for the fingerprint generation. The proposed framework provides a novel solution of legal content distribution, but it does not include collusion resistance and user privacy. Similarly, \cite{lhh10} proposed a P2P system which provides secure distribution of copyright-protected music contents. In this framework, the RSA public-key cryptosystem is used to generate a unique digital fingerprint for every end user within the network. Then, the generated fingerprint is embedded into the music file such that the music provider can establish the identification of any end user performing an unauthorized re-distribution of the file. The proposed system provides a secure mean for distributing large-scale music contents over P2P networks, but it fails to offer privacy to the end users.

Most of the past studies focused on either providing copyright protection to content owners or privacy to end users. Our work differs from existing studies in a way that we focus on the design and implementation of the multimedia content distribution over the P2P network that provides both multimedia security and privacy at a reduced
computational cost to the merchant and the end user.

\section{PSUM model}
\label{sec:PSUM}

This section describes the design and functionality of PSUM. In Section \ref{sec:3.1}, we define the role of each entity and list the notations that are used in the design of PSUM. Section \ref{sec:3.2} defines the functionality requirements and security assumptions. Three different types of attack models are described for PSUM in Section \ref{sec:3.3}. In Sections \ref{sec:3.4}, \ref{sec:3.4.3}, \ref{sec:3.4.4} and \ref{sec:3.5}, we detail the design of PSUM, which includes the fingerprint generation, the base and supplementary files generation and distribution protocols, as well as the traitor-tracing and dispute resolution protocols.

\subsection{System entities and parameters}
\label{sec:3.1}
In PSUM, a hybrid P2P network is opted as a platform for content distribution, since it consumes less network resources and is more scalable than centralized P2P systems. Moreover, the idea of centralized and P2P distribution can easily be achieved by using a hybrid P2P system, since multiple coordinators, called super peers, can easily manage both base file and supplementary file distribution. Also, a public key infrastructure (PKI) is considered for providing a public/private key pair for each entity. Moreover, an offline external certification authority ($CA_{ext}$) is assumed in PSUM for validating the real identity of a buyer by providing a signed public-key certificate to the buyer. It is a one-time process that is executed offline.
In addition to $CA_{ext}$, PSUM involves seven entities and the function of each entity is defined as follows:

\begin{itemize}
	\item A merchant ($M$) is an entity that distributes the copyrighted content to the end users (peers) of PSUM. It is involved in the base file (\textit{BF}) generation and distribution, the supplementary file (\textit{SF}) generation and distribution, traitor tracing and dispute resolution protocols.
	\item A buyer (peer $B_i$) is an entity that can either play a role of a data requester or provider. A buyer is involved in the acquisition of \textit{BF} from the merchant, the distribution of \textit{SF} in PSUM and a dispute resolution if he/she is found guilty of copyright violation.
	\item A super peer (\textit{SP}) (a.k.a. index server) is a reputed peer with additional facilities who is assigned the role of the coordinator for a small portion of the group of peers. Each \textit{SP} maintains a list of the peers connected to the network and acts as a central coordinator. However, \textit{SP} stores peers’ pseudonyms instead of their addresses. The peers send their queries to \textit{SP} for downloading their files of interest. Initially, \textit{SP}s are provided with \textit{SF} from $M$ at the system start-up.
	\item A Certification authority ($CA_R$) is a trusted party that is responsible of issuing certificates to the buyer for the acquisition of \textit{BF} from $M$ and \textit{SF} from other peers. The certificate is used to certify that the pseudonym is correctly registered to $CA_R$ and $CA_R$ knows about the real identity of the buyer. The authentication between different peers is done without involving $CA_R$.
	\item A monitor (\textit{MO}) functions as a trusted party which is responsible for the generation of collusion-resistant fingerprint codes. The existence of \textit{MO} ensures that the generated fingerprints are not revealed to $M$ and the buyer, thus resolving the problems of customer’s rights and non-repudiation. \textit{MO} is also responsible for assigning segments of fingerprint codeword $s_j$ to a set of proxy peers ($Pr_{j}$, for $j = 1, \ldots, n$) in such a way that proxy peers are unable to frame an honest buyer by colluding. In addition, \textit{MO} provides traceability of a buyer by executing a traitor tracing algorithm in case of a piracy claim by $M$. In case of a dispute resolution between $M$, a buyer, and a judge, \textit{MO} provides the pseudonym of the guilty buyer to the judge.
	\item A proxy peer (\textit{Pr}) is responsible for querying content of \textit{BF} available at $M$’s end with the pre-assigned bits of a fingerprint codeword and transferring the retrieved content to the buyer.
	\item A judge ($J$) is assumed to be a trusted party which resolves the disputes between $M$ and a buyer with the cooperation of \textit{MO} and $CA_R$.
	\end{itemize}

Table \ref{tab1} describes the relevant terms and parameters used in PSUM to benefit our readers.

\begin{table}[htp]
	\caption{Parameters and Notations}
	\label{tab1}
	\begin{center}
		\begin{tabular}{>{\centering\arraybackslash}p{4cm}>{\centering\arraybackslash}p{10cm}}	\hline
			\bf{Parameter} & \bf{Specification}\\ \hline
			$B_i$ & $i^{th}$ buyer\\	
			\addlinespace[-1.85mm]
			$P_{B_{i}}$ & Pseudonym of the buyer $B_i$ \\
			\addlinespace[-1.85mm]
			$M$ & Merchant\\
			\addlinespace[-1.85mm]
			\textit{MO} & Monitor\\
			\addlinespace[-1.85mm]
			$Pr_{j}$ & $j^{th}$-Proxy peer\\
			\addlinespace[-1.85mm]
			\textit{SP} & Super peer \\
			\addlinespace[-1.85mm]
			\textit{BF} & Base file \\
			\addlinespace[-1.85mm]
			\textit{SF} & Supplementary file \\
			\addlinespace[-1.85mm]
			$CA_R$ & Certification authority of PSUM\\
			\addlinespace[-1.85mm]
			$CA_{ext}$ & Offline certification authority\\
			\addlinespace[-1.85mm]
			$J$ & Judge \\
			\addlinespace[-1.85mm]
			$a$ & Approximation coefficients\\
			\addlinespace[-1.85mm]
			$a'_{j}$& Permuted and encrypted $a$\\
			\addlinespace[-1.85mm]
			$r$ & Secret shared between $CA_R$ and $B_i$\\
			\addlinespace[-1.85mm]
			$a_3/a_4$& 3/4-level discrete wavelet transform approximation coefficients\\
			\addlinespace[-1.85mm]
			$f_i$& Fingerprint of the buyer $B_i$\\
			\addlinespace[-1.85mm]
			$\mathrm{Cert}_{CA_{R}}(K^{*}_{pB_{i}})$ & Anonymous certificate of $B_i$ certified by $CA_R$ \\
			\addlinespace[-1.85mm]
			$\mathrm{Cert}_{CA_{R}}(M)$ & Certificate of $M$ certified by $CA_{ext}$ \\
			\addlinespace[-1.85mm]
			$Sign_{B_{i}}()$ & Signature of $B_i$ using his/her private key\\
			\addlinespace[-1.85mm]
			$Sign_{P_{B_{i}}}()$ & Signature of $B_i$ using his/her anonymous key\\
			\addlinespace[-1.85mm]
			$n$ & Number of proxy peers\\
			\addlinespace[-1.85mm]
			$s_j$ & Segments of the fingerprint $f_i$\\
			\addlinespace[-1.85mm]
			$ps_j$ &Permuted segments assigned to $Pr_j$\\
			\addlinespace[-1.85mm]
			 $f_{a^0_j}$ & Permuted fragments of $a$ for $0$-bit\\
			 \addlinespace[-1.85mm]
			 $f_{a^1_j}$ & Permuted fragments of $a$ for $1$-bit\\
			 \addlinespace[-1.85mm]
			$\sigma_{j}$ & Set of permutation keys\\
			\addlinespace[-1.85mm]
			$l$ & Length of the permuted fingerprint segment\\
			\addlinespace[-1.85mm]
			$\tau$ & Fixed time period set for \textit{MO}\\
			\addlinespace[-1.85mm]
			$c$ & Number of colluders\\
			\addlinespace[-1.85mm]
			$\in$ & Probability of accusing an innocent end user\\
			\addlinespace[-1.85mm]
			$N$ & Total number of end users in PSUM\\
			\addlinespace[-1.85mm]
			$m$ & Length of a fingerprint code\\
			\addlinespace[-1.85mm]
			$\Delta$ & Quantization constant\\
			\addlinespace[-1.85mm]
			$K_{ses_{j}}$ & Set of one-time session keys\\
			\addlinespace[-1.85mm]
			$X$ & Original content\\
			\addlinespace[-1.85mm]
			$Y$ & Pirated copy\\
			\addlinespace[-1.85mm]
			$pc$ & Pirated codeword\\
			\addlinespace[-1.85mm]
			($K_{P_{M}},K_{S_{M}}$) & Public and private key pair of $M$\\
			\addlinespace[-1.85mm]
			($K_{P_{MO}},K_{S_{MO}}$) & Public and private key pair of \textit{MO}\\
			\addlinespace[-1.85mm]
			($K_{P_{B_{i}}},K_{S_{B_{i}}}$) & Public and private key pair of $B_i$\\
			\hline
 		\end{tabular}
	\end{center}
\end{table}

\subsection{Design and security model of PSUM}
\label{sec:3.2}
In this section, the design requirements and general and security assumptions of PSUM are described.

\subsubsection{Design requirements}
\label{sec:3.2.1}
In the following, the design requirements related to the construction of PSUM are defined:
\begin{itemize}
	\item $M$ should be able to trace and identify an illegal re-distributor in case of finding a pirated copy with the help of \textit{MO}, $J$ and $CA_R$.
	\item $M$ should not be able to frame an honest buyer of an illegal re-distribution.
	\item A malicious buyer who has re-distributed an unauthorized copy should not be able to claim that the copy was created by $M$ or a collusion of $Pr_j$.
	\item The possible collusion of $Pr_j$ should be unable to frame an honest buyer.
	\item The real identity of $B_i$ should remain anonymous during transactions unless he/she is proven guilty of copyright	violation.
	\item The identity of $B_i$ should not be linked to his/her activities such as purchasing, transferring of file and so on.
	\item $J$, with the help of \textit{MO}, should be able to resolve the disputes without involving $B$ in the process.
	\item The reconstruction of the original file from \textit{BF} and \textit{SF} should be performed at the $B$’s end.
	\item \textit{BF} cannot be shared within the end users of PSUM.
\end{itemize}

\subsubsection{General assumptions}
\label{sec:3.2.2}

In this sub-section, the general assumptions related to the construction of PSUM are defined.
\begin{itemize}
	\item At the system start-up, the bootstrapping is carried out via a well-known booting node. The booting node is a known and trusted peer of the network. The selection of the booting node is out of the scope of this paper.
	\item In order to deliver \textit{BF} from $M$ to $B_i$, \textit{MO} selects a fixed number ($n$) of $Pr_j$.
	\item The number of proxy peers $n$ and the length of the fingerprint $m$ are known constants of PSUM.
	\item $Pr_j$ must follow each other in a sequential manner to transfer \textit{BF} to $B_i$ from $M$.
	\item In order to protect data privacy during \textit{BF} exchange, \textit{MO} must wait for some time until at least two buyers request for a content from $M$. This step is enforced on \textit{MO} to ensure that $M$ obtains no knowledge about which coefficient of $a$ is accessed and transferred to $B_i$ ($a$ is an approximation coefficient obtained from the $L$-level DWT).
\end{itemize}

\subsubsection{Security assumptions}
\label{sec:3.2.3}

The security assumptions of PSUM are defined in this sub-section.

\begin{itemize}
	\item $M$ and $B_i$ do not trust each other but they both trust \textit{MO}. Because of the anonymity of the embedding procedure, \textit{MO} generates the collusion-secure fingerprints as this is the only party that is trusted by both $M$ and $B_i$ to generate a valid fingerprint. Also, in case of traitor-tracing process, it is expected that \textit{MO} does not form a coalition with any other party to frame $B_i$.
	\item The fingerprint codes used in PSUM provide a resistance against a given number of colluders ($c = 3$) as specified	by \cite{Nu10} codes.
	\item The embedded fingerprint is imperceptible and robust against common signal processing attacks, and the fingerprint	extraction process is blind.
	\item The permutation keys $\sigma_j$ (for $j = 1, \ldots, n$) are generated by $B_i$ to perform the permutation of a fingerprint codeword to be assigned to $Pr_j$. The purpose of generating $\sigma_j$ is to ensure that a collusion of malicious $Pr_j$ is	unable to generate a valid fingerprint codeword or a fingerprinted content. $\sigma_j$ may be generated by \textit{MO}, instead of $B_i$, but this might increase the overheads of \textit{MO}, which as a result of being overloaded, could potentially become a bottleneck of PSUM. Thus, in order to avoid creating such a situation in PSUM, which might result in performance degradation of the system, the buyers are given the responsibility of generating $\sigma_j$.
	\item $Pr_j$ are not trusted and the content transferred through them is encrypted in such a way that only $M$ and $B_i$ have access to the clear-text.
	\item The real identity of each entity is validated by an external (offline) $CA_{ext}$. Thus, each entity has a public key certificate signed by $CA_{ext}$. $CA_{ext}$ keeps track of all the identities to be sure that they remain unique and also to revoke an identity of a malicious entity.
	\item Each entity ($M$, \textit{MO}, $Pr_j$, $B_i$, $CA_R$, $J$) is supposed to have a public key $K_p$, a private key $K_s$. Public-key cryptography is restricted to the encryption of small-length binary strings such as symmetric session and permutation keys, and data signing.
	\item Before joining PSUM, each peer is authenticated by an internal $CA_R$ of the system. $CA_R$ validates the identity of a peer from $CA_{ext}$. After successful verification, each peer has a private key and a public key certified by	$CA_R$. $CA_R$ generates a random number $r$ and shares it with an authenticated peer for the generation of a pseudoidentity.
	\item PSUM uses hash functions (e.g. based on SHA-1) to generate unforgeable and verifiable pseudo-identities for each entity. The hash is assumed secure and cannot be reversed.
	\item Each buyer can generate multiple pseudo-identities and anonymous certificates depending on his/her anonymity	requirement. For example, if a buyer makes two transactions once a month, then he/she can use a single pseudo-identity
	and an anonymous certificate both times without worrying about linkability of his/her online activities with the pseudo-identity and an anonymous certificate. However, if a buyer purchases content frequently, he/she might require more pseudo-identities and anonymous certificates to preserve his/her anonymity during
	transactions with $M$ and different peers.
\end{itemize}

\subsection{Attack models}
\label{sec:3.3}
This section describes three types of attack models for PSUM: (1) the first attack model is specific to the proposed asymmetric fingerprinting protocol, i.e. it addresses privacy and security attacks on a buyer from malicious entities, (2) the second attack model is related to a general weakness of a digital fingerprinting scheme, i.e. collusion attacks, and (3) the third attack model is watermarking-specific, i.e. it focuses on the common signal processing attacks that are applied onto the marked content either to alter or remove the embedded signal.

\paragraph{\textbf{\textit{1. Privacy and security attacks on a buyer:}}} The following type of attacks are aimed to de-anonymize and accuse an innocent buyer of illegal re-distribution of the purchased content.

\begin{enumerate}[a.]
	\item Different transactions carried out by a buyer with a same pseudo-identity are linkable to one another and	an attacker could infer some private information of a buyer through data mining techniques.
	\item A malicious entity may try to find two different but real identities such that the two identities have the same pseudo-identity. It might then use one of the two identities to impersonate the buyer to obtain a	fingerprinted copy of the content that would be linked to the impersonated buyer.
	\item $M$ and one or more proxy peers may collude to create a new fingerprinted content.
\end{enumerate}

\paragraph{\textbf{\textit{2. Collusion attacks:}}} The collusion attack from a group of malicious buyers (colluders), combining several copies with the same content but different fingerprints to try to remove the embedded fingerprints or frame honest buyers, is the major challenge to digital fingerprinting. If a digital fingerprint is not properly designed, a fingerprinting system might fail to detect the traces of any fingerprints under collusion attacks with only a few colluders. To ensure the reliable tracing of true traitors and avoid framing honest buyers, linear (averaging) and
non-linear (maximum, minimum and median) collusion attacks are performed (details are provided in Section \ref{sec4.1.3}).

\paragraph{\textbf{\textit{3. Watermarking attacks:}}} An attack succeeds in defeating a watermarking scheme if it impairs the fingerprint beyond acceptable limits while maintaining the perceptual quality of the attacked data. In PSUM, we have selected
state-of-the-art audio watermarking \citep{Wang2014} and video watermarking \citep{Leelavathy2011} schemes as building blocks for embedding the collusion-resistant fingerprinting code in both audio and video contents. The selected embedding schemes provide excellent robustness and transparency results against common signal processing attacks (details are provided in Section \ref{sec5}).

Formal proofs and an informal security analysis for the first attack model, “privacy and security attacks on a buyer” are presented in Sections \ref{sec4.1.1} and \ref{sec4.1.2}, respectively. The security analysis for the ``collusion attack" model is discussed in Section \ref{sec4.1.3}. Finally, the evaluation of the robustness of the embedded fingerprint against the ``watermarking attacks" model is discussed in Section \ref{sec5}.

\subsection{Overview of PSUM}
\label{sec:3.4}
The different sub-protocols of PSUM are detailed in the following sections.

\subsubsection{Fingerprint generation}
\label{sec:3.4.1}
The fingerprint $f_i$ is generated by \textit{MO} using the \cite{Nu10} codes algorithm. The fingerprint generation algorithm takes $\in$, $N$ and $c$ as inputs, and outputs a collection $F = (f:1,\ldots,f_N)$ of binary codewords ($f_i$) of size $m$ and a secret bias vector $p$. The length of the fingerprint is calculated as $\in_0 = Ne^{-\alpha_{0}m}$, where the value of $\alpha_{0}$ is $0.0725$. The details of the algorithm can be found in \cite{Nu10}.

\subsubsection{Base file generation and distribution protocol}
\label{sec:3.4.2}

\textit{BF} is designed to have a small size and is distributed from $M$ to the end users of PSUM on receiving a payment. The DWT is applied on a multimedia content to split it into approximation and detail coefficients. The approximation coefficients are then split into second-level approximation and detail coefficients, and the process is repeated as many times as desired (levels of decomposition). PSUM supports both audio and video files.

\begin{figure}[H]
	\centering
	\subfloat[Set of segments $s_j$ of a fingerprint $f_i$]{ 
			\includegraphics[width=12cm]{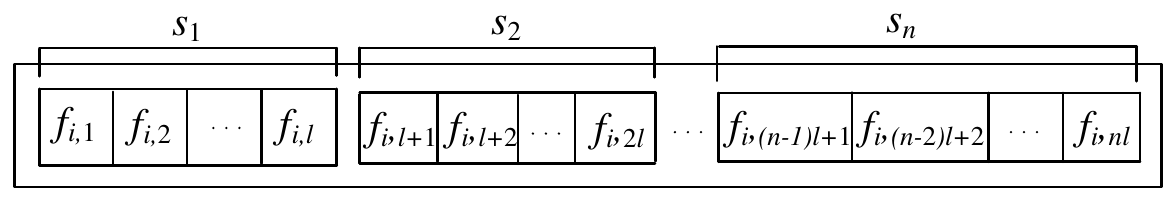}
		\label{fig:fig1}}\\
	\subfloat[Sets of contiguous fragments ($f_{a^0_j},f_{a^1_j}$) of two variants of $a$]{
			\includegraphics[width=12cm]{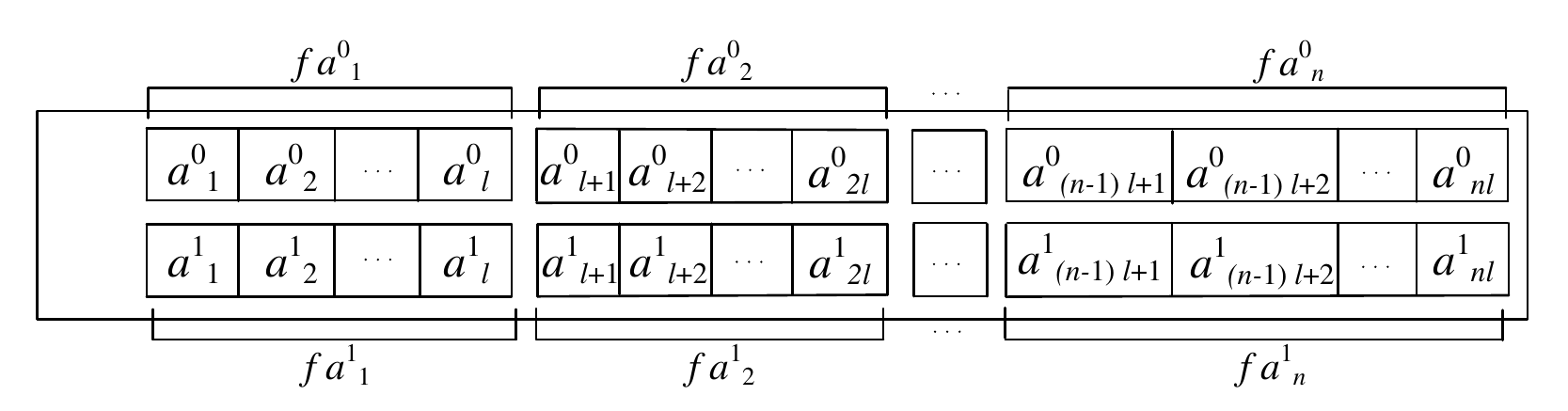}
		\label{fig:fig2}}\\
		\caption{Fingerprint segments and set of fragments of pre-computed $a$ assigned to $n$ proxy peers}
	\label{fig:segments}
\end{figure}

In case of an audio file, the DWT decomposition results in approximation ($a$) and detail ($d$) coefficients. We have considered level-$4$ DWT decomposition for audio to obtain a convenient trade-off between the robustness, capacity and transparency properties of watermarking. The $4$-level approximation coefficients, $a_4$, are then used twice to imperceptibly embed a fingerprint of all ones and all zeros, using a blind, robust and secure QIM-based watermarking technique \cite{xwzxy13}, as shown in Fig. \ref{fig:fig2}. The two variants of $a'_4$ form \textit{BF} in a binary form, which is then distributed to a buyer through a protocol providing asymmetric fingerprinting (see Protocol 1).

For generation of a video \textit{BF}, the first task is to extract the significant frames (intra-frames or I-frames) from a video file. The intra-frames (I-frames) are coded without reference to other frames (inter-frames). The inter-frames (P and B frames) use pseudo-differences from the previous and the next frame and, hence, these frames depend on each other. It is not advisable to embed data both into intra and inter-frames. Thus, we have used only I-frames which contain the most significant information. We have used the Canny Edge Difference technique \citep{kc13} to obtain the I-frames. In the Canny Edge Difference Detection method, a difference between two consecutive frames is calculated and if this difference exceeds a calculated threshold value, we obtain a key frame. The extracted key frames are converted from the RGB format to Y'UV, whereas the remaining frames, i.e. P and B-frames are saved in an original video format. The Y'UV model defines a color space in terms of one luminance (Y') and two chrominance (UV) components. For each I-frame, we choose the Y' component and apply $4$-level DWT to obtain $a$ and $d$. Two variants of $a'_4$ (one embedded with all zeros and the other one embedded with all ones) are obtained in a binary form by using a blind QIM-based watermarking technique \citep{Leelavathy2011} similar to the audio \textit{BF} creation process.

\begin{figure}[ht]
	\centering
	\includegraphics[width=14cm]{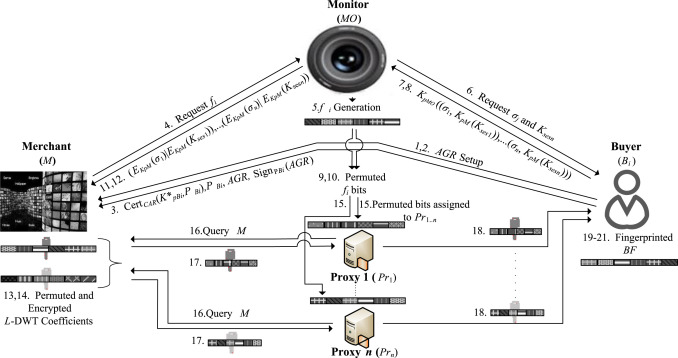}
	\caption{\textit{BF} distribution protocol}
	\label{fig:fig12}
\end{figure}

On receiving a file request from a buyer $B_i$, \textit{SP} provides him/her the details of $M$ who has the requested content. For a secure distribution of \textit{BF} to the buyer, $M$, \textit{MO}, $B_i$ and a selected set of $Pr_j$ perform an asymmetric fingerprinting protocol. Fig. \ref{fig:fig12} illustrates the distribution protocol of \textit{BF}.

In case of an audio file, a buyer receives BF in a binary (``dat") format and \textit{SF} in a ``wav" format. We take an example of an audio file ``Hugewav.wav" (whose details are provided in Table 3) to explain the creation of \textit{BF} and \textit{SF}. ``Hugewav.wav" is a two-channel stereo audio file with each channel sampled $44,100$ times per second with $16$ bits per sample. Level-$4$ DWT is applied to both the channels separately in order to obtain level-$4$ approximation coefficients (yielding $19,300$ coefficients for each channel). The approximation coefficients of both channels are saved in a double-precision floating-point format that occupies 8 bytes in computer memory. Thus, the size of the ``dat" file containing the approximation coefficients of both the channels corresponds to $19300\times2\times8\approx 0.30$ MB. The coefficients are saved in a lossless compressed format to produce \textit{BF} of size equal to $0.29$ MB, which is approximately $10$ times smaller than original ``Hugewav.wav" file. For \textit{SF}, an inverse $4$-level DWT is performed on the detail coefficients (of both the channels) with the approximation coefficients set to zero. \textit{SF} is saved in an original audio format, i.e. ``wav" using $32$ bits per sample in order to preserve the details, yielding a final file with size equal to $5.94$ MB. The size of \textit{SF} is thus, the double of the size of the original audio file due to the fact that \textit{SF} is formed with double-bit precision values.

\begin{figure}[ht]
	\centering
	\includegraphics[width=12cm]{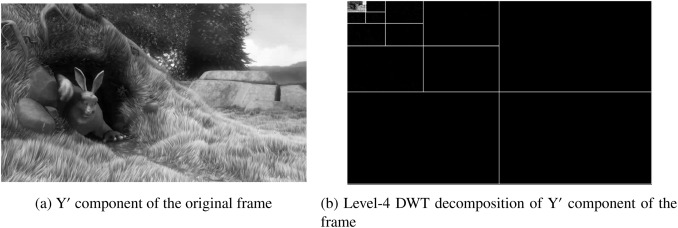}
	\caption{Level-$4$ DWT approximation coefficients and an original key frame of ``Dragon.avi"}
	\label{fig:fig3}
\end{figure}

For video files, \textit{BF}s received by the buyers are in binary (``dat") format, whereas \textit{SF}s are in ZIP form. An example of the DWT decomposition of an I-frame (for a sample public domain video) is provided in Fig. \ref{fig:fig3}. Fig. 3a shows the Y' component of the original frame, whereas Fig. 3b illustrates the level-$4$ decomposition of the Y' component of the same I-frame. Only the coefficients at the top-left area ($LL_4$) are taken for \textit{BF}.

As video is regarded, we take, as an example, the ``Dragon.avi" (whose details are provided in Table 4) to explain the creation of \textit{BF} and \textit{SF}. ``Dragon.avi" contains a total of $32,975$ frames, of which, $2228$ are I-frames (or key frames) and $30,747$ are P and B-frames. Similar to the case of the audio \textit{BF}, the level-4 approximation coefficients ($a_4$) are saved in a double-precision floating-point format. The size of the I-frames for this video is $320\times240$ pixels, whereas the $a_4$ coefficients are a matrix with $20\times15$ values. Hence, the size of $a_4$ coefficients of $2228$ I-frames, saved in ``dat" format, corresponds to $20\times15\times8\times2228$ bytes $\approx 5$ MB. These $a_4$ coefficients and the index of the I-frames are stored in a lossless compressed binary format (``dat") to produce a \textit{BF} of size $4.80$ MB. Thus, the size of \textit{BF} is approximately $10$ times shorter than the original file size of ``Dragon.avi", that is equal to $51.10$ MB. In contrast to \textit{BF}, the \textit{SF} contains the P and B-frames, the audio tracks of the original video file, the level-4 detail coefficients of the Y' component of the key frames, and the UV components of I-frames. An inverse level-4 DWT is applied on these detail coefficients with the approximation coefficients set to zero to build a ``reconstructed" Y' component, which is combined with the UV components to obtain $2228$ frames in RGB format. Then, these ``reconstructed" I-frames and the inter-frames are used to generate a video file (``avi"), in which the ``reconstructed" $2228$ frames are inserted into the same positions as that of the I-frames in the original video. Thus, the \textit{SF} consists of $32,975$ frames saved in a video format along with the audio tracks of the file. The contents of \textit{SF} are compressed using lossless data compression method (e.g. deflate) to produce \textit{SF} in ZIP form with a size equal to $69.40$ MB.

Note that, in both cases (audio and video), the most relevant data are stored in \textit{BF}, whereas \textit{SF} is absolutely useless without the corresponding \textit{BF}. For audio, \textit{SF} has all the relevant low frequency coefficients set to zero, and thus, it only contains the high frequency data that, uncombined with the low frequency counterpart (stored only in \textit{BF}), does not produce any usable sound signal. Similarly, for video, the P and B-frames are useless without their corresponding I-frames. In addition, the I-frames of \textit{SF} only contain the UV components of the original key frames, which, again, are useless without the corresponding Y' component, and the (high frequency) detail coefficients of Y'. Hence, \textit{SF} cannot be used to reconstruct a valid image for the key frames without the approximation coefficients of Y' that are stored only in \textit{BF}. Without a usable Y', the ``reconstructed" key frames of \textit{SF} do not produce a usable video.

In short, for both audio and video files, \textit{BF} stores the most significant low-frequency coefficients of the DWT decomposition of the relevant signal or component and not only headers or metadata.

\begin{algorithm}[H]
	\floatname{algorithm}{\textit{Protocol}}
	\caption{\textit{\textbf{(Steps that are executed between \textit{MO}, $M$, $B_i$, and proxy peers to distribute fingerprinted \textit{BF} to $B_i$)}}}
	\label{protocol1}  
	\begin{algorithmic}[1]
		\item Before starting a purchase negotiation of the multimedia content with the merchant, $B_i$ generates a pseudo-identity to keep his/her anonymity. For pseudo-identity generation, $CA_R$ generates a random number $r$ and shares it only with $B_i$.
		\item $B_i$ negotiates with $M$ to set-up an agreement ($AGR$) that explicitly states the rights and obligations of both parties and specifies the multimedia content ($X$). $AGR$ uniquely binds this particular transaction to $X$. During the negotiation process, $B_i$ uses his/her pseudonym $P_{B_{i}}$ to keep his/her anonymity.
		\item After the negotiation, $B_i$ generates a key pair ($K^{*}_{pB_{i}},K^{*}_{sB_{i}} $), signs the public key with his/her private key, and sends $Sign_{B_{i}}(K^{*}_{pB_{i}}, P_{B_{i}})$ source to $CA_R$. $CA_R$ verifies $Sign_{B_{i}}(K^{*}_{pB_{i}}, P_{B_{i}})$ using the public key of $B_i$. If valid, he/she generates an anonymous certificate $\mathrm{Cert}_{CA_{R}}(K^{*}_{pB_{i}}, P_{B_{i}})$ and sends it to $B_i$. $B_i$ then sends $\mathrm{Cert}_{CA_{R}}(K^{*}_{pB_{i}}, P_{B_{i}})$, $Sign_{P_{B_{i}}}(AGR)$, $AGR$ and $P_{B_{i}}$ to $M$.
		\item $M$ verifies the received certificate using the public key of $CA_R$ and the signature of the agreement using the certified key $K^{*}_{pB_{i}}$. If the received data is valid, then $M$ generates a transaction ID ($TID$) for keeping a record of the transaction between him/her and $B_i$, and sends a request for a fingerprint to \textit{MO} by sending $\mathrm{Cert}_{CA_{R}}(K^{*}_{pB_{i}}, P_{B_{i}})$, $\mathrm{Cert}_{CA_{R}}(M)$, $TID$, $AGR$, $P_{B_{i}}$ and $Sign_{P_{B_{i}}}(AGR)$. If the received certificates and signatures are not valid, then the transaction is terminated by $M$.
		\item \textit{MO} validates the certificates and signatures of $M$ and $B_i$ from $CA_R$. After successful verification, \textit{MO} generates a 
		Nuida's $c$-secure codeword $f_i$ of length $m$ and randomly selects $n$ proxy peers ($Pr_j$, for $j=1,\ldots,n$) for a secure transfer of fingerprinted \textit{BF} from $M$ to $B_i$.
		\item \textit{MO} sends a request for permutation keys $\sigma_j$ and session keys $K_{ses_{j}}$ to $B_i$.
		\item After receiving the request from $M$, $B_i$ generates $n\footnotemark{}$ random $\sigma_j$ (for $j=1,\ldots,n$) of length \footnotemark{} $l=\left\lfloor m/n \right\rfloor$ and $n$ session keys $K_{ses_{j}}$. $K_{ses_{j}}$ are generated to be shared with $M$, such that $Pr_j$ that are responsible for transferring the fingerprinted $a$ to $B_i$ are unable to see the clear-text of $a$.
		\item $B_i$ encrypts $K_{ses_{j}}$ with $K_{p_{M}}$ and sends $K_{p_{MO}}(\sigma_j,K_{p_{M}}(K_{ses_{j}}))$ to \textit{MO}.
		 \item \textit{MO} decrypts $K_{p_{MO}}(\sigma_j,K_{p_{M}}(K_{ses_{j}}))$ with $K_{s_{MO}}$ and obtains $\sigma_j$ and $K_{p_{M}}(K_{ses_{j}})$.
		 \item \textit{MO} divides $f_i$ into $n$ segments ($s_j$) of length $l$ (as shown in Fig. 1a) and permutes each segment using the permutation keys $\sigma_j$ in the same order as received by $B_i$.
		 \item \textit{MO} waits for a specific time $\tau$ such that it receives multiple requests of a content from different buyers. If by that specified time \textit{MO} receives other requests, then the steps 1-10 are repeated for the new buyer.
		 \algstore{myalg}
	 \end{algorithmic}
 \end{algorithm}
\footnotetext[1]{$n$ is a known constant of the system.}
\footnotetext[2]{The length $l$ is equal to $\left\lfloor m/n \right\rfloor$ for all the permutation keys $\sigma$ except for maybe the last one, which maybe shorter.}

\begin{algorithm}[H]
	\begin{algorithmic}[1]
		\algrestore{myalg}
		\item For each buyer, \textit{MO} sends $E_{K_{p_{M}}}(\sigma_j)|E_{K_{p_{M}}}(K_{ses_{j}})$ to the corresponding $M$.
		\item $M$ decrypts $E_{K_{p_{M}}}(\sigma_j)|E_{K_{p_{M}}}(K_{ses_{j}})$ with $K_{s_{M}}$ and obtains $\sigma_j$ and $K_{ses_{j}}$.
		\item $M$ permutes sequentially both pre-computed variants of $a$ with $\sigma_j$. An exchange of $\sigma_j$ between $M$ and \textit{MO} is performed to ensure that proxy peers do not get the positions of the permuted fingerprint bits. $M$ then encrypts the permuted a variants with $K_{ses_{j}}$.
		\item The contiguous permuted fingerprint segments ($ps_j$) are then sequentially assigned to $n$ proxy peers by \textit{MO}.
		\item $Pr_j$ contacts $M$ in a sequential manner to obtain the fragments of encrypted and permuted approximation coefficients $\left\{f_{a^0_j},f_{a^1_j}  \right\}$.
		\item $M$ sends a set of encrypted and permuted fragments of pre-computed approximation coefficients $\left\{f_{a^0_j},f_{a^1_j}\right\}$ to $Pr_j$.
		\item $Pr_j$ selects the correct pre-computed (encrypted and permuted) approximation coefficients $a_j$ from the received coefficients $\left\{f_{a^0_j},f_{a^1_j}\right\}$ using the assigned permuted fingerprint segment $ps_j$, as shown in Fig. \ref{fig:fig4}.
		\item When $B_i$ receives $\left\{f_{a^0_j},f_{a^1_j}\right\}$ from $Pr_j$, he/she permutes back the encrypted coefficients with $\sigma^{-1}_{j}$. With $K_{ses_{j}}$, $B_i$ decrypts the received encrypted approximation coefficients and obtains the fingerprinted coefficients of \textit{BF}.
		\item $B_i$ obtains his/her complete copy of \textit{BF} by composing all the coefficients received sequentially from all $Pr_j$.
		\item An inverse level-$L$ DWT is applied on \textit{BF} to get a fingerprinted \textit{BF}, which is then recombined with \textit{SF} obtained from the P2P network.
			\end{algorithmic}  
	\end{algorithm}

The security analysis of Protocol \ref{protocol1} is provided in Section \ref{sec4.1}.

\subsection{Supplementary file generation and distribution}
\label{sec:3.4.3}

For an audio content, an inverse level-3/4 DWT is performed on the detail coefficients to obtain \textit{SF} in ``wav" form. Other formats, such as binary and text, can also be used. In case of a video file, the inverse level-4 DWT of detail coefficients for I-frames, the inter-frames (P and B) and the audio of the content constitutes \textit{SF} in compressed (ZIP) form. The distribution of \textit{SF} is carried out in P2P fashion.

On joining the system, a peer constructs an onion-path with existing peers which points to it and adds this path to his/her associated \textit{SP}. By doing so, a requesting peer (\textit{RP}) can use this onion-path to contact the content providing (\textit{CP}) peer while knowing nothing about the \textit{CP}'s identity.

The peer requests for a particular file to \textit{SP} of his/her group. If found, it displays the list of the peers having that particular file; else it sends a request for the file to other connected \textit{SP}s. The other \textit{SP}s, on finding the particular \textit{CP}, send the response to the requesting \textit{SP}. \textit{SP} then establishes a path between \textit{RP} and that \textit{CP} peer. After receiving a positive reply from the \textit{CP} peer, the requesting peer initiates a two-party authenticated key exchange (AKE) protocol \citep{wang2012} to authenticate each other identities and exchange the content of \textit{SF} anonymously. For secure exchange of data, a one-time session key is generated during the AKE protocol to encrypt the content of \textit{SF}. The details of \textit{SF} distribution can be found in \cite{qmr15}.

\begin{figure}[ht]
	\centering
	\includegraphics[width=12cm]{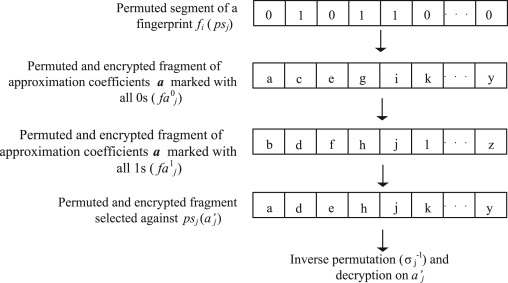}
	\caption{Fragment selection}
	\label{fig:fig4}
\end{figure}

\subsection{Traitor tracing}
\label{sec:3.4.4}

Once a pirate copy $Y$ of content $X$ is found, $M$ extracts the fingerprint by decomposing the pirated content Y with the same wavelet basis used in the fingerprint insertion step. This gives the approximation coefficient matrix in which the pirated code $pc \in \left\{0,1\right\}^{*}$ is embedded. The watermark detection technique \citep{Leelavathy2011,xwzxy13} is applied on the approximation coefficient matrix to extract $pc$. Then $M$ sends the extracted $pc$ to \textit{MO} which performs the tracing algorithm of Nuida's codes to identify the colluder(s). The details of the tracing algorithm can be found in \cite{Nu10}. The real identity of the buyer is not known to \textit{MO}, only the pseudo-identity of the guilty buyer is revealed.

\subsection{Arbitration and identification protocol}
\label{sec:3.5}

The goal of the arbitration and identification protocol, performed between $M$, \textit{MO}, $CA_R$ and $J$, is to reveal the real identity of the traitor or reject the claims made by $M$. In order to reveal the real identity of the traitor, \textit{MO} sends ($Y$, $pc$, $K_{p_{MO}}(f))$ and $M$ sends $\mathrm{Cert}_{CA_{R}}(K^{*}_{pB_{i}}, P_{B_{i}})$, $AGR$, $Sign_{P_{B_{i}}}(AGR)$ and $K^{*}_{pB_{i}}$ to $J$. $J$ verifies the validity of all the certificates and the signatures. If valid, it asks \textit{MO} to decrypt $E_{K_{p_{MO}}}(f)$. If $pc$ and $f$ match with a high correlation, it requests $CA_R$ to give the real identity of the buyer. Otherwise, the buyer is proved innocent.

\subsection{Comparison with \cite{qmr15}}
\label{sec:3.6}

Although the suggested PSUM scheme and the previous work published by \cite{qmr15} share some ideas, there are several significant differences between these two schemes. Both systems are based in partitioning the multimedia file into a small-sized base file and a large-sized supplementary file to lessen the computational cost of the merchant. Since the supplementary file is useless without the base file, it can be distributed in P2P fashion without many security constraints.

The differences between both methods are mainly related to the creation and distribution of the small-sized base file. In the system of \cite{qmr15}, the base file was distributed in a centralized manner from the merchant to each buyer. To obtain buyer frameproofness, this required that the merchant encrypted a few selected coefficients of the base file using homomorphic encryption, embedded the fingerprint in the encrypted domain, and transmitted the encrypted-fingerprinted file to the buyer. Since encrypting the coefficients expands the data significantly, only a few of them were actually used to embed the fingerprint. The remaining coefficients were encrypted together in block form to avoid further data expansion. On receiving the base file, the buyer should decrypt it using his/her private key and recombine it with the supplementary file to obtain the final fingerprinted content. Even though the homomorphic encryption is performed only to the small-sized base file, this implied computational burden both for the merchant (for encryption and embedding) and for the buyer (for decryption).

On the other hand, the proposed PSUM system completely avoids homomorphic encryption. Here, each coefficient of the base file can be used for embedding a bit of the fingerprint with no additional cost, since the two possible embeddings (``1" or ``0") for each coefficient are pre-computed. Public-key encryption of the base file is avoided by fragmenting the base file and sending the fragments to a group of proxies. The proxies receive the fragments of the base file and the fingerprint permuted such that they cannot collude to obtain a valid fingerprint. In the buyer's side, public-key decryption is not necessary. Once the fragments are received, the permutation can be reversed by the buyer, who applies symmetric decryption to reconstruct the base file, which can be finally combined with the supplementary file to obtain the fingerprinted content.

These relevant differences have a remarkable effect on the efficiency of the system, as shown by means of several experiments in Section \ref{sec5}. Besides, PSUM provides better collusion resistance as shown in Section \ref{sec4.1.3} (last row of Table 2). The round-off errors that occur in the system of \cite{qmr15}, caused by the use of homomorphic encryption, are completely avoided in the proposed PSUM scheme. In addition, more coefficients are used for embedding the fingerprint in PSUM compared to \cite{qmr15}. This leads to a more effective application of the fingerprinting codes and, hence, to better collusion resistance.

\section{PSUM analysis}
\label{sec4}

In this section, we provide an analysis of PSUM in terms of privacy and security. The security analysis provide formal proofs and informal analysis concerning the correctness and soundness of Protocol \ref{protocol1}. We also discuss practicality issues in PSUM due to deployment of trusted third parties ($CA_R$, $J$ and \textit{MO}).

\subsection{Security analysis}
\label{sec4.1}

This section analyzes the privacy and security properties of PSUM according to the design requirements and the attack models presented in Secs \ref{sec:3.2} and \ref{sec:3.3}, respectively.

\subsubsection{Formal analysis of Protocol 1}
\label{sec4.1.1}

Formal proofs are provided in this section to analyze the security of Protocol \ref{protocol1}.

\begin{thm}
	A malicious buyer with a pseudonym $P_E$, impersonating a legitimate buyer $B_i$ to initiate a purchase protocol with $M$ and later obtain a fingerprinted copy in order to frame $B_i$ for illegal re-distribution, can be detected in Protocol \ref{protocol1}.
\end{thm}
\begin{pf}
	In PSUM, the anonymity of a buyer's identity is obtained using a one-way cryptographic hash function $h$. This hash function provides a pseudo ID which can be used for anonymous authentication and communication. An attempt of de-anonymization attack by a malicious buyer is withstood by the collision resistance of the hash function, i.e. it is computationally infeasible to find a pair ($x,y$) such that $h(x)=h(y)$. Moreover, for a hash function with $w$-bit hash values, $2^{w/2}$ calculations are required to find a collision with probability $1/2$, which is infeasible for $w\geq128$. In our design, we have considered SHA-1 with $w=160$ bits for high security such that it is computationally infeasible for an attacker to compute $2^{80}$ calculations to find a real identity from a pseudo ID. Furthermore, a malicious buyer cannot use the pseudo ID of other buyer because he/she does not know the secret number $r$ shared by the buyer with $CA_R$. Thus, $P_E$ would be detected during the verification phase of Protocol \ref{protocol1}.
	\end{pf}

\begin{thm}
	In Protocol \ref{protocol1}, a malicious proxy peer $E$ is unable to obtain a secret permutation key $\sigma_j$ transmitted from $B_i$ to \textit{MO} or from \textit{MO} to $M$.
\end{thm}

\begin{pf}
\textit{MO} initiates a fingerprinting protocol with $M$ and $B_i$ only after verification of certificates and signatures from $CA_R$. The secret permutation key transferred between $B_i$ and \textit{MO} or between \textit{MO} and $M$ is encrypted with the public key of \textit{MO} or $M$, respectively. Thus, in order to obtain $\sigma_j$, $E$ needs the private key of \textit{MO} or $M$ to decrypt $K_{p_{MO}}(\sigma_j)$ or $K_{p_{M}}(\sigma_j)$.
	\end{pf}

\begin{thm}
	An honest buyer is protected, in Protocol \ref{protocol1}, from a conspiracy attack of malicious proxy peers who try to recombine their segments of a fingerprint and/or the fingerprinted content obtained from the merchant.
\end{thm}

\begin{pf}
	In case $Pr_j$ try to obtain a correct fingerprint by recombining their assigned permuted segments $ps_j$ (with length of each segment equal to $l$), $Pr_j$ would need to compute $l!$ combinations each on the colluded fingerprint $f'_i$. Thus, with more $m$-bits in $f_i$, $Pr_j$ would need to do increased number of permutations in order to obtain a correct fingerprint, which would be computationally infeasible.
	
	In the second case, if all $Pr_j$ combine their permuted and encrypted fragments $E_{K_{ses_{j}}}(a'_j)$ obtained from $M$, they cannot decrypt these fragments. 
	The fragments can only be decrypted by $K_{ses_{j}}$, which are known only to $M$ and $B_i$. Hence, $Pr_j$ are unable to obtain clear-text fingerprinted fragments to produce a fingerprinted copy similar to the buyer's copy.
	
	For example, we consider a randomly permuted fingerprint $f_i$ of length $90$-bits and three proxy peers $Pr_1$, $Pr_2$ and $Pr_3$. Each proxy peer carries $30$-bits. In case $Pr_1$, $Pr_2$ and $Pr_3$ collude together and obtain $f'_i$, they need to compute $30!$ combinations each, resulting in $30!\cdot30!\cdot30!=(30!)^{3}$ total combinations to try to get a valid $f_i$.
\end{pf}

\subsubsection{Privacy and security attacks}
\label{sec4.1.2}

In this section, possible privacy and security attacks on a buyer from malicious entities in Protocol \ref{protocol1} are discussed.

\begin{itemize}
	\item \textbf{Buyer frameproofness:} The possible collusion of proxy peers $Pr_j$ cannot frame an honest buyer and held him/her responsible for illegal re-distribution (formally proved in Theorem 3). Also, $M$ alone is unable to produce a fingerprint $f_i$, since \textit{MO} is responsible for generation of $f_i$. However, it may be possible that a malicious $M$ colludes with \textit{MO} to frame an honest buyer for illegal re-distribution. Similarly, another possible collusion can occur between the proxy peers and $M$. In the first scenario, the collusion can be disregarded since \textit{MO} is an entity that is trusted by both $M$ and $B_i$ (as described in Section \ref{sec:3.2.3}). In the second case, when $Pr_j$ query $M$ to obtain the permuted pre-computed $\left\{f_{a^0_j},f_{a^1_j}  \right\}$, it might be possible that both $M$ and $Pr_j$ collude to obtain a valid fingerprint codeword or a fingerprinted copy. Since $M$ has a clear-text of $\sigma_j$, it could permute the fingerprint bits obtained from all the proxy peers by using $\sigma_j$ and obtain a valid fingerprint of a buyer. However, this conspiracy attack against an honest buyer requires that all the proxy peers ($n$) collude with $M$, thus making a collusion size equal to $n+1$. In addition, the merchant would not be interested in forming such a big collusion at a price of being possibly caught, since it is possible that one of the proxy peers be honest and refuse to become a part of this coalition. Then this proxy peer may report about the collusion between $M$ and the remaining proxy peers to \textit{MO}. It may be noted that if less than $n$ proxy peers collude with $M$, then the probability of framing an honest buyer is very low. For example, if $n=10$ with each proxy peer carrying $l=10$ bits and $20$\% of $n$ colludes with $M$, then the probability of obtaining a valid fingerprint is $0.2^{10}\approx 10^{−7}$, which is very low.
	
	In PSUM, this conspiracy attack can be countered by compelling \textit{MO} to wait for a particular time period $\tau$, so that by the expiry of $\tau$, it receives more fingerprint requests from $M$ for different buyers. By doing so, $M$ would be accessed by various $Pr_j$ at a time and keeping record of various bits of multiple proxy peers could be infeasible. Also, a reward mechanism can be introduced within PSUM so that proxy peers can get rewards, such as discounts or bonus points, for their good reputation and reliability. 
	
	Furthermore, it could be possible that $M$ tried to find an identity of the buyer by relating proxies to each buyer. For example, if the permuted and encrypted approximation coefficients are transferred from $M$ to two buyers $B_1$ and $B_2$ through $n$ and $n - 2$ proxy peers, respectively, it would be easier for $M$ to figure out that a particular set of proxy peers $Pr_j$ with $j=1,\ldots,n-2$ are carrying a fingerprint for the buyer $B_2$ or $Pr^{'}_{j}$ with $j=1,\ldots,n$) are carrying $f_i$ for $B_1$. Thus, to avoid a possible attack of $M$ on $B_i$, the number of proxy peers is fixed to $n$.
	
	\item \textbf{Merchant security:} From the perspective of $M$, PSUM is secure because a buyer $B_i$ has no idea about the original digital content and the embedded fingerprint in the purchased copy. Also, $B_i$ cannot claim that a pirated copy is created by $M$ since the fingerprint is generated by \textit{MO} which is trusted by both $B_i$ and $M$. Thus, $B_i$ cannot accuse \textit{MO} of collaborating with $M$ to frame him/her (as described in Section \ref{sec:3.2.3}). However, there can be two cases where copyright protection scheme could be broken:

\begin{enumerate}
	\item Since the proxies receive the permuted-encrypted coefficients $a^{'}_{j}$, a possible collusion of $B_i$ and all (or some of) $Pr_j$ makes it possible to obtain the complete (or partial) set of coefficients and produce non-fingerprinted copies of the content, as $B_i$ has everything he/she needs, namely, the symmetric key and the permutation keys. In this case, a possible $B_i$ and $Pr_j$ collusion is prevented by assigning the task of selecting $Pr_j$ to \textit{MO}. Consequently, $B_i$ should create a collusion with $Pr_j$ that are anonymous to him/her. But it is too risky, since honest $Pr_j$ would accuse $B_i$ of this misbehavior. However, if it is considered that the risk of this collusion cannot be overlooked (because even a single fragment leaked could be dangerous), there is a solution. The communication between $Pr_j$ and $B_i$ could be implemented using a path created by \textit{MO}. In this way, the buyer would not even know the $Pr_j$ who originated the fragment and he/she would be required to build a collusion with all the nodes of the all the paths for all the fragments, which is unrealizable.
	\item Malicious $Pr_j$ may choose a combination of approximation coefficients that does not correspond to the fingerprint bits. For example, $Pr_j$ may choose the $1$-coefficient when the corresponding bit is $0$. In this scenario, the malicious $Pr_j$ would not obtain any benefit by acting in this way, since the content obtained by $B_i$ would not carry a valid fingerprint. However, this malicious act could be evaded, again, by using the paths created by \textit{MO} between $Pr_j$ and $B_i$. Some of the nodes of this path could randomly decide to send the fragment to \textit{MO} to check whether the embedded information coincides with the corresponding fingerprint segment. In case of a mismatch, $Pr_j$ would be detected as malicious. Thus, it would be risky for $Pr_j$ to act in this way, since they would not know the nodes of the path created by \textit{MO}.
\end{enumerate}

\item[] Furthermore, from an analysis of $B_i$ frameproofness property, it is obvious that there is a very low probability that a correct fingerprint or a fingerprinted content is obtained from a possible collusion between the proxy peers and $M$. Thus, it is impossible for $B_i$ to deny an act of copyright violation. Also, PSUM provides a tracing mechanism to unambiguously identify a copyright violator once a pirated copy $Y$ is found.
 
\item \textbf{Buyer's privacy:} Although anonymous certificates provide anonymity to $B_i$, the transactions carried out by the same pseudo ID can be linked to one another by a global or a semi-global adversary that continuously monitors multiple points of a network. Data mining involves the use of sophisticated data analysis tools to discover previously unknown, valid patterns and relationships in large data sets \citep{az96,ed99}. Therefore, the use of these techniques makes it possible for an adversary to analyze $B_i$'s online transactions and infer some sensitive personal information (e.g. pseudo-ID, IP address, etc.), thus posing a threat  to $B_i$ \citep{wa08}. For example, aggregation, one of the data mining techniques, can be used by an attacker to combine distinguishable pieces of information from different transactions over a certain time period, and associate the observed attributes to uncover the identity of the targeted buyer with high confidence \citep{Ir11}. The solution to this problem is to allow a buyer to apply for multiple pseudonyms and anonymous certificates simultaneously and randomly choose one for each transaction. Alternatively, a privacy-preserving data mining method such as mapping can be used in which the buyer׳s pseudo ID can be converted to another value, which then provides identification on the other end. However, the countermeasures to defeat data mining, other than the use of multiple pseudonyms for the same user, are out of the scope of this work.
 
\item \textbf{Man-in-the-middle attack:} 
\end{itemize} 
In PSUM, the deployment of PKI ensures mutual authentication between entities ($M$, $B_i$, and \textit{MO}), and thus the communication between the entities is authenticated and the possibility of eavesdropping can be defied. Furthermore, secret keys transferred from $B_i$ to \textit{MO} or from \textit{MO} to $M$ are encrypted with the receivers public keys to prevent tampering of the secret data.

\subsubsection{Collusion attacks}
\label{sec4.1.3}

This section presents the robustness of the fingerprint against the linear (averaging) and non-linear (minimum, maximum and median) collusion attacks presented in Section \ref{sec:3.3}. The attacks are performed on a sample video file ``Dragon.avi" (details of ``Dragon" video file are provided in Table 2) with varying number of colluders $U$. Under the averaging attack, each pixel in the colluded video is the average of the corresponding pixels of the fingerprinted videos associated with the colluders $U$. For the minimum, maximum and median attacks, each pixel in the colluded video is the minimum, maximum or median, of the corresponding pixels of the fingerprinted video.

In order to evaluate the robustness of the fingerprint against more than three colluders, we have selected \cite{nfhkwoi07} collusion-secure ($c$-secure) codes, since the collusion-resistant fingerprinting codes used in PSUM \citep{Nu10} provides security against three colluders only. A fingerprinting code is called $c$-secure, if it is secure against collusion attacks by up to $c$ pirates, and is equipped with a tracing algorithm, which can output a colluder (or a set of colluders) correctly with an overwhelming probability (as described in Section \ref{sec:coll}). The tracing algorithm of \cite{nfhkwoi07} first calculates a score ($S_i)^{(j)} \in R$ for the $j$-th bit of the $i$-th buyer by a certain function, which depends on discrete bias distribution, and then calculates the total score $S_i$ of the $i$-th buyer as follows:

\begin{equation*}
S_i = \sum_{j=1}^{m} (S_i)^{j}.
\end{equation*}

The adopted \cite{nfhkwoi07}'s $c$-secure codes provide resistance against $c$   colluders with $c\leq c_0$ ($c_0$ is the coalition size to be resisted). The collusion resistance (at least $c_0=5$ colluders are traced) of Nuida et al.'s fingerprinting codewords under linear (average) and non-linear (minimum and maximum) collusion attacks is presented in Table 2.

Table 2 shows the number of colluders $U$ which have been successfully traced through the tracing algorithm of Nuida et al.'s codes. In all cases, the colluders have been successfully traced by analyzing a colluded video copy $Y$. In order to test the resistance of the fingerprint against more than $3$ colluders, the fingerprint codewords are generated using $c=4$ and $c=5$ in \cite{nfhkwoi07}'s codes, which results into codewords with an increased length $m$. Here, for the evaluation of collusion resistance of the fingerprint, we have restricted the number of colluders $U$ to $5$ due to a fact that an increase in $U$ degrades the quality of the content (due to an increased length $m$). Thus, to provide a better trade-off between the collusion resistance property and imperceptibility, a lower value of $c_0$ is selected.

Table 2 also shows the comparative analysis of video file ``Dragon.avi" with a system proposed by \cite{qmr15} in terms of collusion resistance. It can be seen, from the table, that all the colluders $U$ are successfully traced in PSUM for different attacks. In the system proposed by \cite{qmr15}, most of the colluders are traced except in one case ($U=5$). For the case $U=5$, when maximum and minimum attacks are applied to the fingerprinted copy, $4$ out of $5$ colluders are successfully identified in the system of \cite{qmr15} in comparison to PSUM, where all $5$ colluders are successfully traced.

\begin{figure}[ht]
	\centering
	\includegraphics[width=13cm]{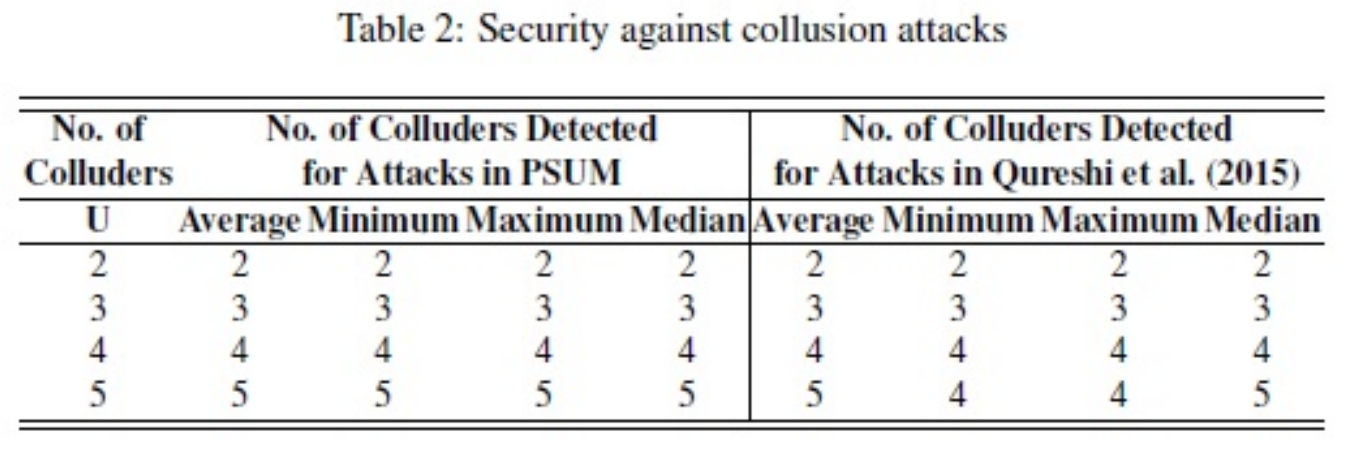}
	\label{fig:fig5}
\end{figure}

\subsection{Deployment of trusted third parties}
\label{sec4.2}

In the presence of trusted third parties, the performance of PSUM is only affected if they are used in every phase of the system. In PSUM, there are three trusted parties ($J$, $CA_R$ and \textit{MO}), and their roles are limited to one or two processes. In this section, we discuss the practicality issues related to the deployment of these parties in PSUM.

\begin{itemize}
\item \textbf{Internal certification authority:} In order to satisfy the anonymity requirement of the secure content distribution protocol, as stated in Section \ref{sec:3.2.1}, a buyer's real identity must remain anonymous to the merchant during the transaction except when he/she is found guilty of illegal re-distribution. An appropriate way of providing revocable anonymity is to use pseudo-identity and an anonymous key pair validated by a certification authority. If there is no certification authority in PSUM, and the buyers use self-generated pseudo-identities, then there would be no way of tracing a malicious buyer since each buyer could use multiple pseudo-identities and can even impersonate other buyers. Thus, there is always a trade-off between anonymity and accountability. Increased anonymity can cause problems in the identification of a copyright violator which in turn could be a problem for a merchant. Thus, to ensure accountability and revocable anonymity in PSUM, the presence of the internal certification authority ($CA_R$) is worth it. Its role is to generate one-time public certificate and a secret number $r$, which only take a few milliseconds of time. $CA_R$ also validates the anonymous key pair used by the authenticated buyer during the anonymous content distribution protocol, which again takes only a few milliseconds of time. $CA_R$ must remain online in PSUM, since it is required in a secure content distribution protocol. Since $CA_R$ needs to store the records of information of all the buyers of PSUM, it must have storage capacity. Considering the ever-decreasing prices of storage devices, the overhead introduced by $CA_R$ is negligible in running a system that provides copyright protection, privacy and accountability simultaneously.
\item \textbf{Judge:} The judge $J$ is a trusted third party that is not involved in any other protocol of PSUM except the arbitration and identification protocol. The presence of $J$ in PSUM ensures that $B_i$ does not need to participate in the dispute resolution protocol, and the identity of $B_i$ is not exposed until he/she is found guilty of illegal re-distribution. Hence, $J$ is only called in case $M$ finds a pirated copy, thus he/she does not need to be online all the time. Furthermore, $J$ is a memoryless trusted third party, since there is no need for $J$ to store any kind of information related to $B_i$ or any other party involved in the arbitration and identification protocol.	
\item \textbf{Monitor:} A monitor \textit{MO} is a trusted party used to provide framing resistance to $B_i$ from $M$ in the \textit{BF} distribution protocol. If \textit{MO} is not considered in a \textit{BF} distribution protocol, then $M$ is solely responsible for generation and embedding a fingerprint into the content requested by $B_i$. However, this creates a customer's right problem. Similarly, if $B_i$ generates his/her unique fingerprint and sends it securely to $M$ for embedding into the content, it causes a repudiation issue, since a guilty $B_i$ producing unauthorized copies could be able to repudiate the fact and claim that these copies were possibly made by $M$. In case both $M$ and $B_i$ generate their own fingerprint, and the jointly computed fingerprint is embedded into the content by $M$, this creates a problem of quality degradation or ambiguity attacks. Therefore, the existence of \textit{MO} ensures that the fingerprint embedded into the content is not revealed to either $M$ or $B_i$. It is proven by \cite{pg99} that efficient fair exchange protocols cannot be completely fair without the help of a trusted third party that is mutually trusted by both $M$ and $B_i$ performing the protocol. Thus using \textit{MO} is a price worth paying if it can turn a \textit{BF} distribution protocol into a practical alternative. \textit{MO} is not involved in the embedding operation; it is used in generation of a collusion-resistant fingerprint, segmentation and permutation of the fingerprint, and communication with $M$, $B_i$ and $Pr_j$ in the \textit{BF} distribution protocol. The overhead costs of \textit{MO} in the \textit{BF} distribution protocol is provided in Section \ref{sec5.3}.
	\end{itemize}

\section{Experimental results}
\label{sec5}

In this section, the experimental results of the fingerprint embedding scheme against watermarking attacks are presented in terms of imperceptibility and robustness. Furthermore, the performance evaluation of the protocols of PSUM is provided regarding computational and communication costs, and cryptographic overhead.

Five experiments, including the computation of transparency to show the objective difference grade (ODG) and peak signal-to-noise ratio (PSNR) of the fingerprinted audio and video files, the evaluation of the robustness of the fingerprint against common signal processing attacks, the calculation of the computational and response time and the calculation of the cryptographic overhead, have been performed to show the performance of PSUM. The experiments have been developed in Matlab and C++ with six audio and six video files, with varying sizes, on a workstation equipped with an Intel i-7 processor at $3.4$ GHz and $8$ GB of RAM. The fingerprint generation, file partitioning, \textit{BF} distribution and traitor-tracing protocols are implemented in Matlab, whereas the \textit{SF} distribution protocol is executed in the C++ programming language.

The details of audio and video files are presented in Table 3 and Table 4. Table 3 and Table 4 also present the sizes of \textit{BF} and \textit{SF}, and it can be seen that the size of \textit{BF} are relatively small. For audio files, the experiments are performed for each channel of audio signals separately. Also, the audio \textit{SF} is formed with double-bit precision values since Matlab $7.0$ stores signals as double-precision values and, otherwise, the file reconstruction at the user end would not be perfect due to quantization errors.

\begin{figure}[ht]
	\centering
	\includegraphics[width=14cm]{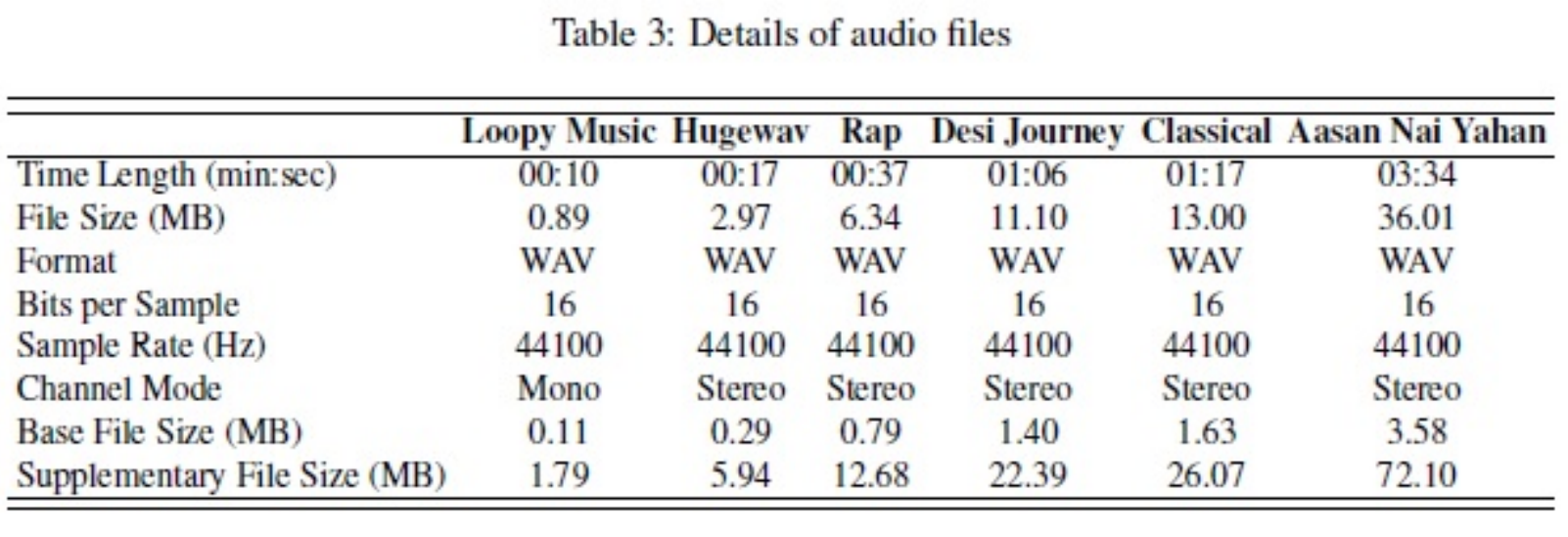}
	\label{fig:fig6}
\end{figure}

\begin{figure}[ht]
	\centering
	\includegraphics[width=14cm]{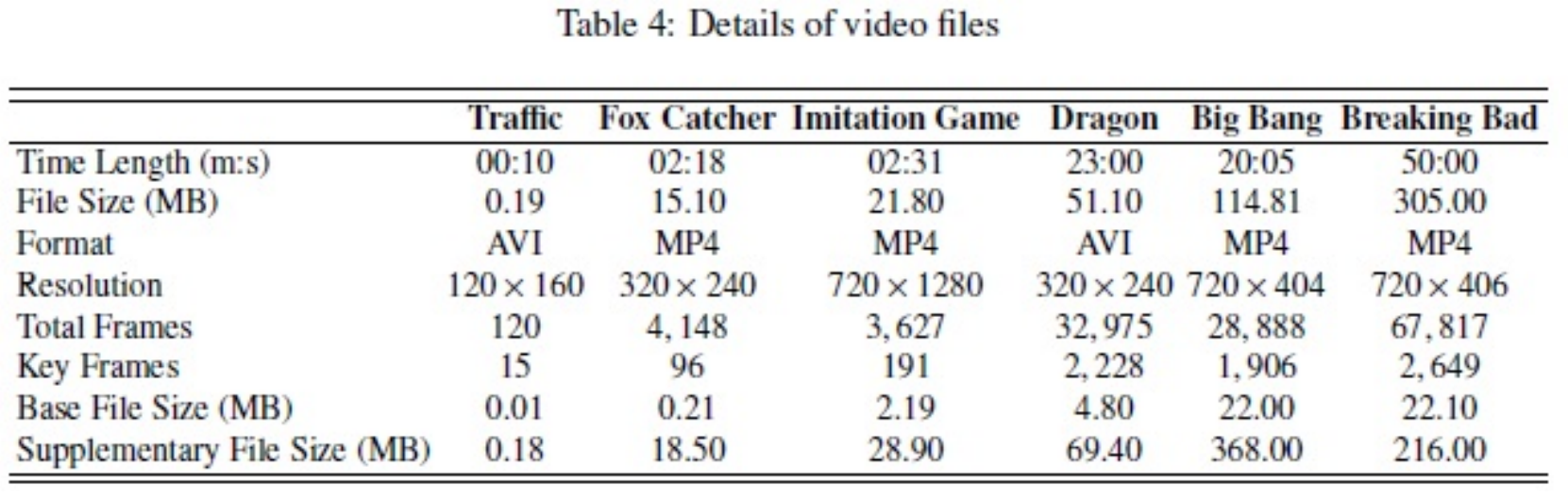}
	\label{fig:fig7}
\end{figure}

\subsection{Transparency of fingerprinted audio and video files}
\label{sec5.1.1}

The ODG is a measurement of audio distortion and is assumed to provide an accurate model of the subjective difference grade results. The ODG results are provided using the advanced perceptual evaluation of audio quality (PEAQ) ITU-R B.S.1387 standard \citep{ttbssbc00} as implemented in the \cite{Opticom} software. Table 5 presents the imperceptibility results as ODG of two fingerprinted audio files ``LoopyMusic.wav" and ``Hugewav.wav" for varying $\Delta$ values, where $\Delta$ is a user-defined positive real number called quantization step. The computed range of the ODG values vary with different $\Delta$ values. For example, in case of ``LoopyMusic.wav", the ODG values are in the range [$−0.15,−2.53$]. Similarly, for ``Hugewav.wav", the computed ODG values are in the range [$−0.01,−0.41$]. This variation in the ODG values depends on $\Delta$ values. We have selected $\Delta=0.25$ for embedding the fingerprint into the remaining four audio files to obtain a convenient trade-off between transparency and robustness. Lower $\Delta$ values would produce better quality audio files, but at a cost of decreased robustness and vice versa.

In Table 6, we present the ODG results of the remaining four audio files (for $\Delta=0.25$). In all cases, the ODG values are between $0$ (not perceptible) and $−1.0$ (not annoying), showing excellent behavior in terms of imperceptibility.

\begin{figure}[ht]
	\centering
	\includegraphics[width=14cm]{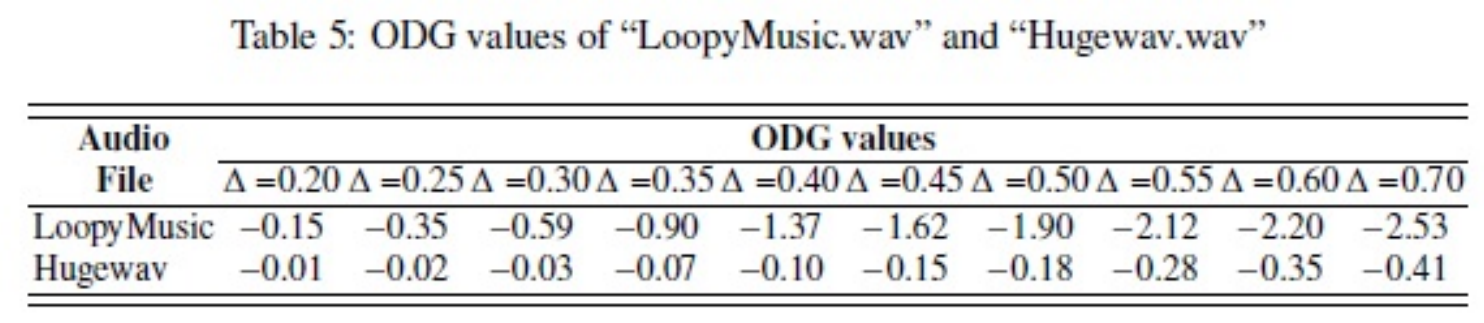}
	\label{fig:fig8}
\end{figure}

\begin{figure}[ht]
	\centering
	\includegraphics{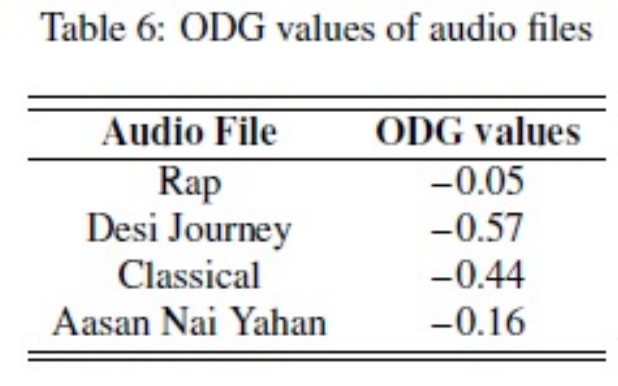}
	\label{fig:fig9}
\end{figure}

For video files, the quality is determined by the PSNR of the fingerprinted video. The PSNR provides a reliable indication of the variation of subjective video quality in decibels (dB). The PSNR values are obtained by using the Moscow State University (MSU) software \citep{MSU}. Similar to audio results, Table 7 presents the imperceptibility results as PSNR of two fingerprinted video files ``Traffic.avi" and ``Imitation Game.mp4" for varying $\Delta$ values. The computed range of the PSNR values vary with different $\Delta$ values. For example, for ``Traffic.avi" and ``Imitation Game.mp4", the PSNR values are in the range [$13.60,42.60$] and [$13.70,67.40$] dB. This variation in the PSNR values depends on $\Delta$ values. We have selected $\Delta=1.10$ for embedding the fingerprint into the remaining four video files to obtain a convenient trade-off between the transparency and robustness properties.

\begin{figure}[ht]
	\centering
	\includegraphics[width=14cm]{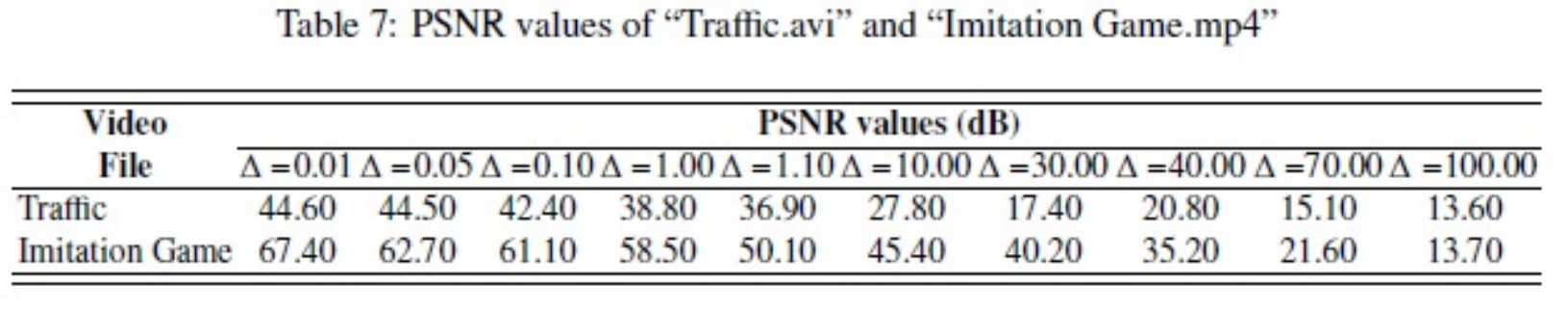}
	\label{fig:fig10}
\end{figure}

Table 8 presents the imperceptibility results of the remaining four video files (for $\Delta=1.10$). It is evident, from Table 8, that the PSNR values are above $35$ dB in each case, and thus it can be inferred that the embedded fingerprint has no perceptible effect on the quality of the video file.

\begin{figure}[ht]
	\centering
	\includegraphics{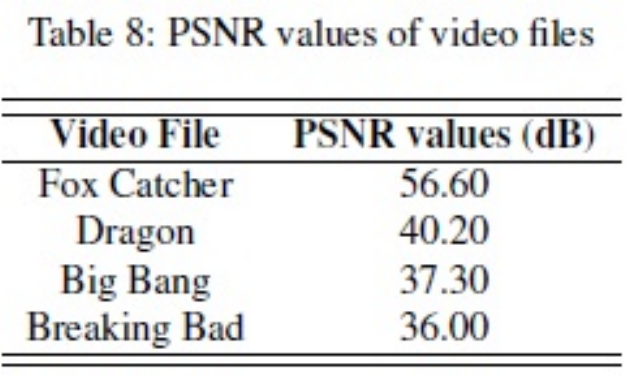}
	\label{fig:fig11}
\end{figure}

In Table 9, we compare the transparency values of ``Loopy-Music.wav" and ``Rap.wav" with other audio watermarking schemes. The state-of-the-art DWT-based audio watermarking schemes \cite{bhse08,Wang2014} have been selected for the comparison purpose. The reason for selecting DWT-based audio watermarking schemes is the fact that PSUM utilizes the idea of partitioning a multimedia file into a base and a supplementary file using the DWT. In order to compare the imperceptibility of the PPSUM's audio watermarking algorithm with the selected state-of-the-art algorithms, the same $\Delta$ value is selected as used in the PSUM's algorithm ($\Delta=0.25$). It can be seen that the embedding algorithm used in PSUM (\cite{xwzxy13}) shows excellent performance compared to the other algorithms in terms of imperceptibility.

\begin{figure}[ht]
	\centering
	\includegraphics[width=14cm]{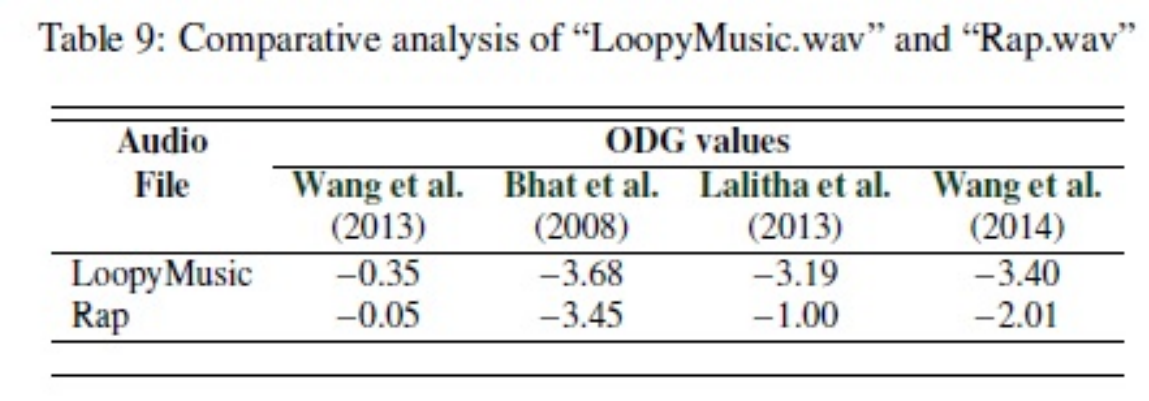}
	\label{fig:fig13}
\end{figure}

Table 10 presents the comparative analysis of the PSNR of two video files, ``Traffic.avi" and ``FoxCatcher.mp4" with other watermarking schemes. The criteria for the selection of the state-of-the art DWT-based video/image watermarking schemes \citep{SPS11,Sh13,Kala13} is similar to selection of the audio watermarking schemes. For comparative purposes, the same $\Delta$ value is selected as used in the PSUM's algorithm ($\Delta=1.10$). It is evident from the table that the embedding algorithm used in PSUM \citep{Leelavathy2011} is more transparent than the compared algorithms.

\begin{figure}[ht]
	\centering
	\includegraphics[width=14cm]{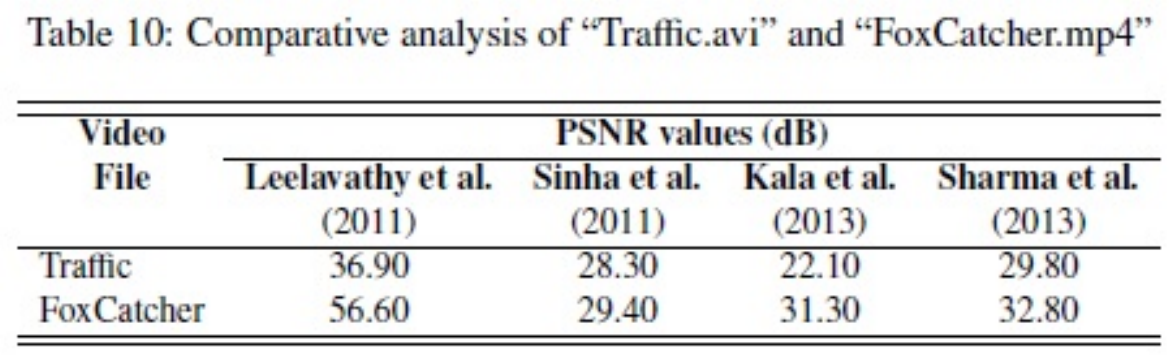}
	\label{fig:fig14}
\end{figure}

\subsection{Robustness of fingerprinted audio and video files}
\label{sec5.2}

Table 11 and Table 12 present the robustness results for two audio files, namely ``LoopyMusic.wav" and ``Rap.wav", against signal processing attacks such as re-quantization, re-sampling, additive white Gaussian noise (AWGN), amplitude scaling, low and high pass filtering, echo addition and MP3 compression. The bit error rate (BER) and normalized correlation (NC) are used to evaluate the robustness between the original fingerprint and the extracted fingerprint. The BER is defined as follows:

\begin{equation*}
BER=\frac{\sum_{j=1}^{m}f_{ij} \oplus f'_{ij}}{m},
\end{equation*}

where $\oplus$ denotes the exclusive OR operation between the original fingerprint $f_i$ and the extracted fingerprint $f'_i$, respectively, $i$ is an index of the buyer and $m$ is the size of the fingerprint code.

NC is defined as follows:

\begin{equation*}
BER=\frac{\sum_{j=1}^{m}f_{ij}f'_{ij}}{\sqrt{\sum_{j=1}^{m}f^{2}_{ij}}\sqrt{\sum_{j=1}^{m}f'^{2}_{ij}}}.
\end{equation*}

If NC is close to $1$, then the similarity between $f_i$ and $f'_i$ is very high. If NC is close to $0$, then the similarity between $f_i$ and $f'_i$ is very low.

The BER values closer to $0$ and NC values closer to $1$ indicate robustness against signal processing attacks. Moreover, the traceability of the fingerprinted copies of ``LoopyMusic.wav" and ``Rap.wav" is evaluated against the signal processing attacks by using the tracing algorithm of \cite{Nu10}. For each attacked audio file, the fingerprint of a buyer is traceable.

In Table 11 and Table 12 , we also present the comparative analysis of ``LoopyMusic.wav" and ``Rap.wav" with the selected audio watermarking algorithms \citep{bhse08,lara13,Wang2014} in terms of BER and NC. It can be seen, from both tables, that the audio embedding algorithm of PSUM \citep{xwzxy13} shows better performance than the compared algorithms in terms of robustness. The results presented in Table 11 and Table 12 show a superior behavior of PSUM's algorithm against AWGN, amplitude scaling, filtering and MP3 compression in comparison with the selected algorithms.

In Table 13 and Table 14, we present the robustness of two video files, namely ``Traffic.avi" and ``FoxCatcher.mp4", against signal processing attacks such as re-scaling, median filtering, blurring, low-pass filtering, AWGN, frame rotation and H.264 compression.

\begin{figure}[ht]
	\centering
	\includegraphics[width=14cm]{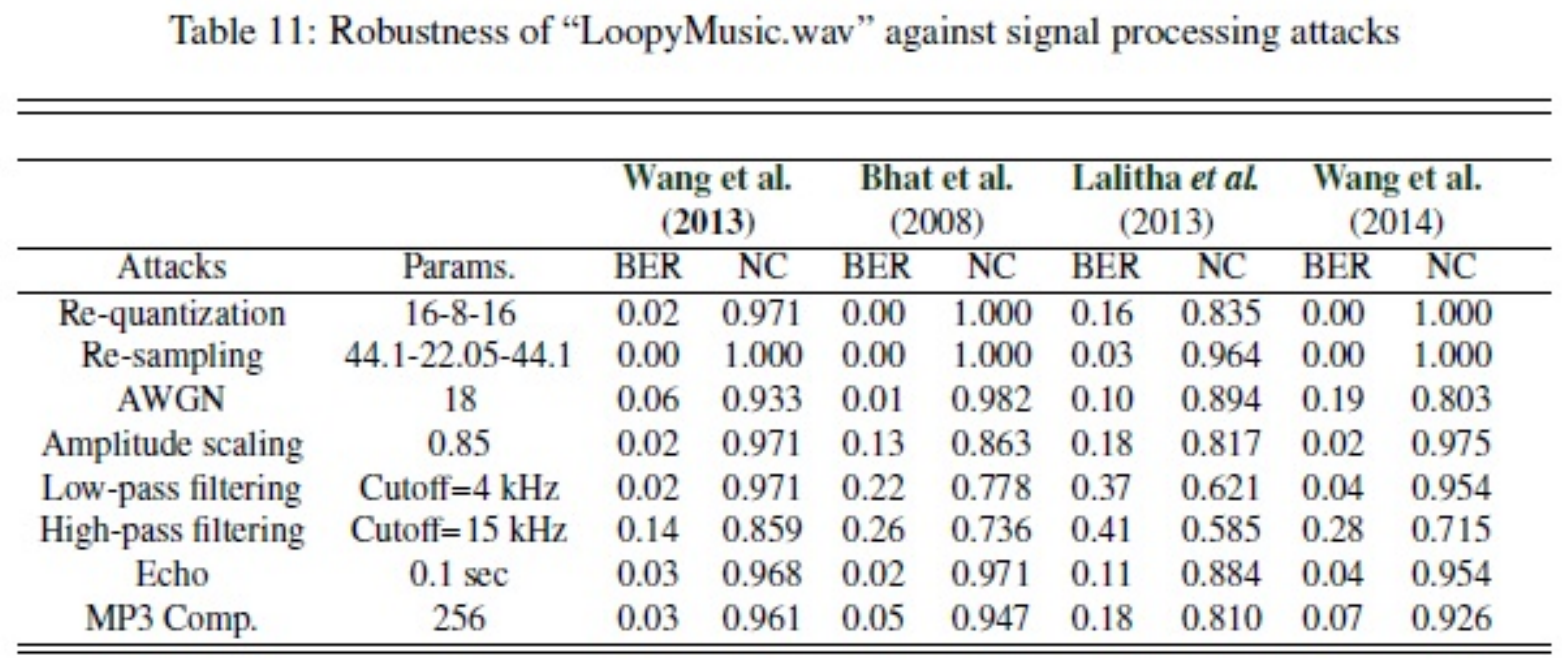}
	\label{fig:fig15}
\end{figure}

The BER and NC are used to evaluate the robustness between the original fingerprint and the extracted fingerprint. Similar to the audio files, we have evaluated the traceability of ``Traffic.avi" and ``Fox Catcher.mp4" fingerprinted video copies against the signal processing attacks by using the tracing algorithm of \cite{Nu10} codes. The fingerprint of a buyer is traceable in each case.

Table 13 and Table 14 also present the comparative analysis of ``Traffic.avi" and ``FoxCatcher.mp4" with the selected video watermarking algorithms \citep{SPS11,Sh13,Kala13} in terms of BER and NC. From the results in both tables, it is evident that the video embedding algorithm of PSUM \citep{Leelavathy2011} provides superior performance against median filtering, low-pass filtering, rotation and H.264 compression attacks than the compared algorithms. Thus, the selected fingerprint embedding algorithm satisfies the fingerprint's robustness requirement.

\begin{figure}[ht]
	\centering
	\includegraphics[width=14cm]{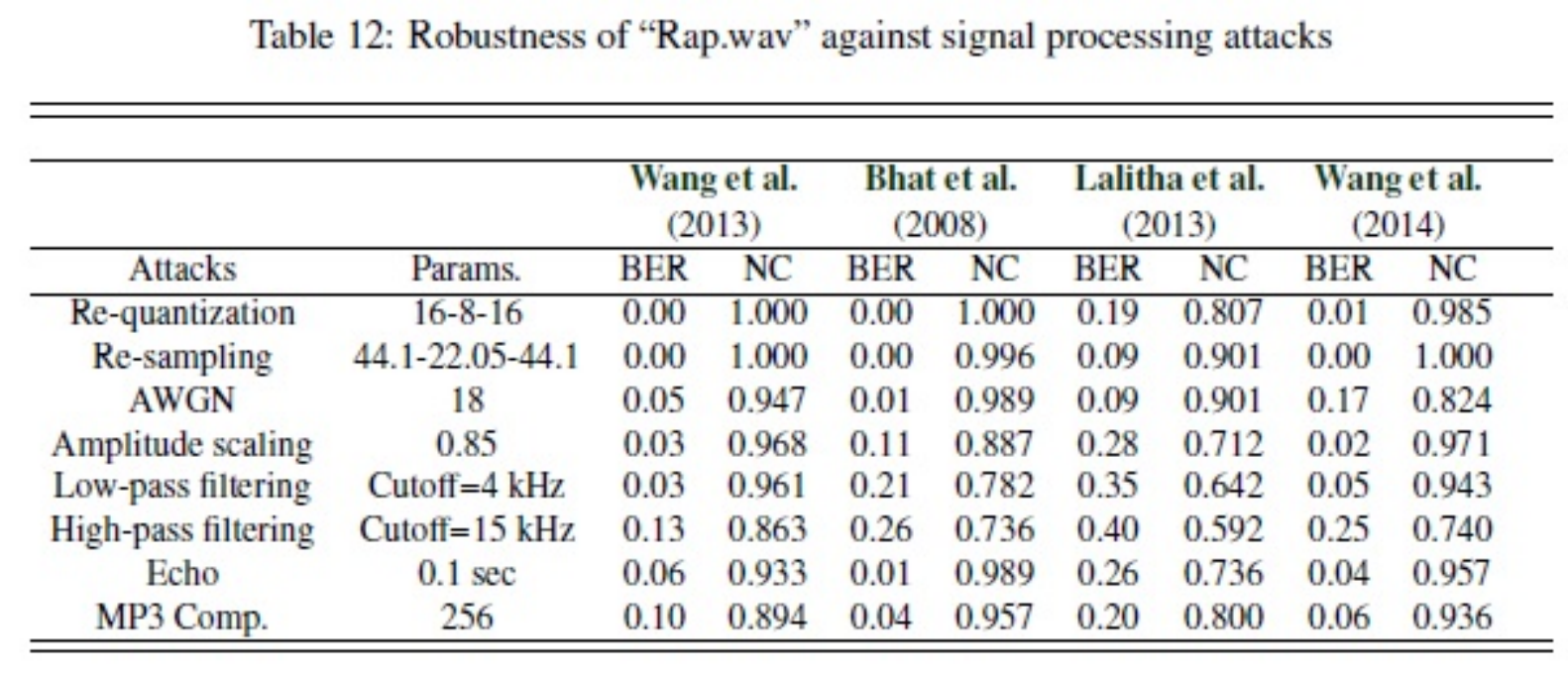}
	\label{fig:fig16}
\end{figure}

\begin{figure}[ht]
	\centering
	\includegraphics[width=14cm]{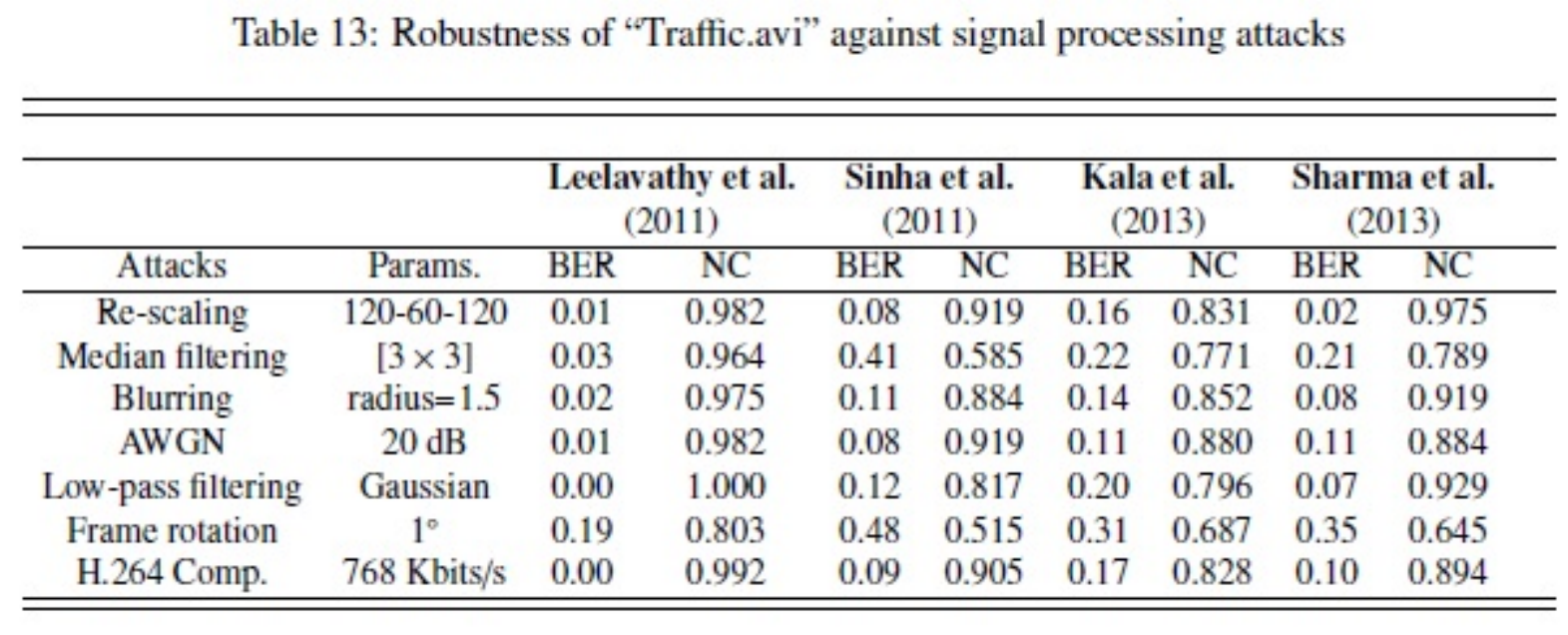}
	\label{fig:fig17}
\end{figure}

\begin{figure}[ht]
	\centering
	\includegraphics[width=14cm]{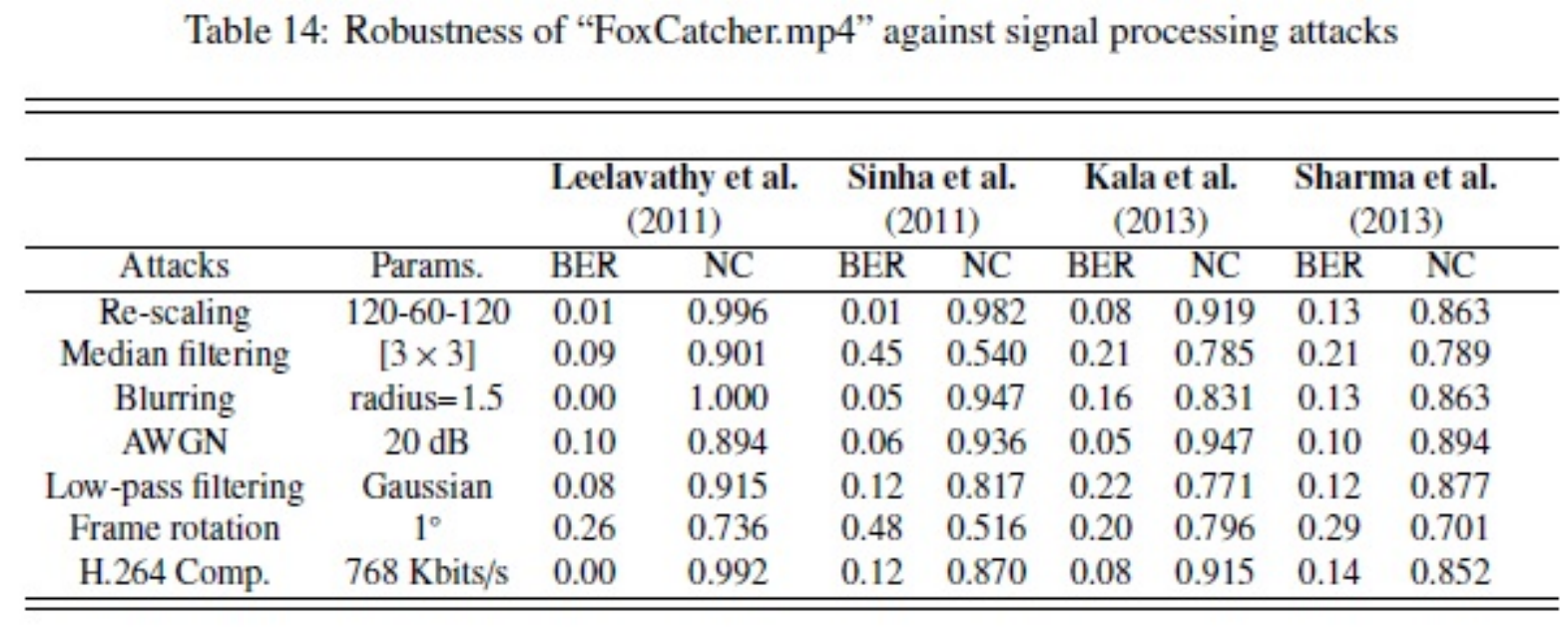}
	\label{fig:fig18}
\end{figure}

\subsection{Computational time}
\label{sec5.3}

In this section the performance of the \textit{BF} distribution protocol for two audio and two video files is discussed in terms of computational time. The overheads are calculated for each party ($M$, \textit{MO}, $B_i$ and $Pr_j$) involved in the \textit{BF} distribution protocol. We have calculated the costs for a scenario where a file has been requested from $M$ by a single buyer.

Table 15 presents the overhead of $M$ for two audio files ``LoopyMusic.wav" and ``Hugewav.wav", and two video files ``Traffic.avi" and ``Breaking Bad.mp4". In the \textit{BF} distribution protocol for an audio file, $M$ is responsible for creating \textit{BF} using DWT, decryption of the permutation ($\sigma_j$) and session ($K_{ses_{j}}$) keys obtained in an encrypted form from \textit{MO}, and permutation and encryption of the pre-computed approximation coefficients.

\begin{figure}[ht]
	\centering
	\includegraphics[width=14cm]{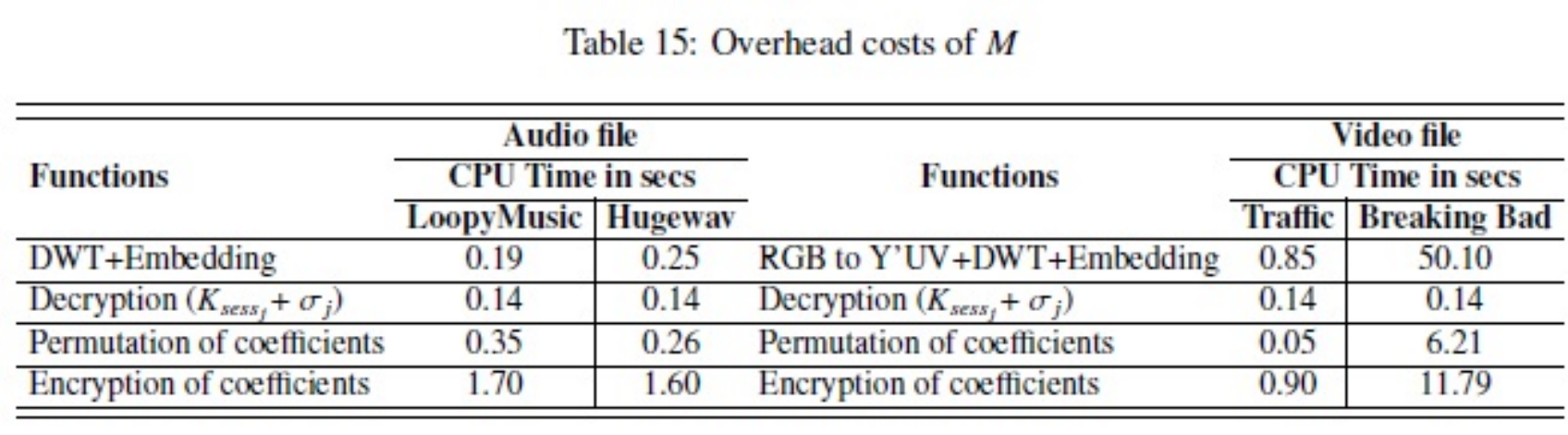}
	\label{fig:fig19}
\end{figure}

In the audio file partitioning algorithm (Section \ref{sec:3.4.2}), the DWT is applied only once to the content to obtain the approximation and detail coefficients. $M$ stores the approximation and detail coefficients of each file, and thus avoids the costs of applying the DWT every time an audio file is requested by a buyer. Similarly, the RGB conversion to Y'UV format and DWT on the luminance (Y') components of the key frames are applied once by $M$ to obtain the approximation and detail coefficients. $M$ stores the approximation and detail coefficients of each video file. By doing so, $M$ is able to avoid the cost of performing RGB to Y'UV conversion and DWT every time a video file is requested by $B_i$. The decryption of permutation and session keys process takes a few seconds and is common for both audio and video files. From Table 15, it can be seen that it takes $M$ a few seconds to encrypt and permute the approximation coefficients of each file, thus the overhead cost of $M$ is trivial in the \textit{BF} distribution protocol.

\begin{figure}[ht]
	\centering
	\includegraphics[width=14cm]{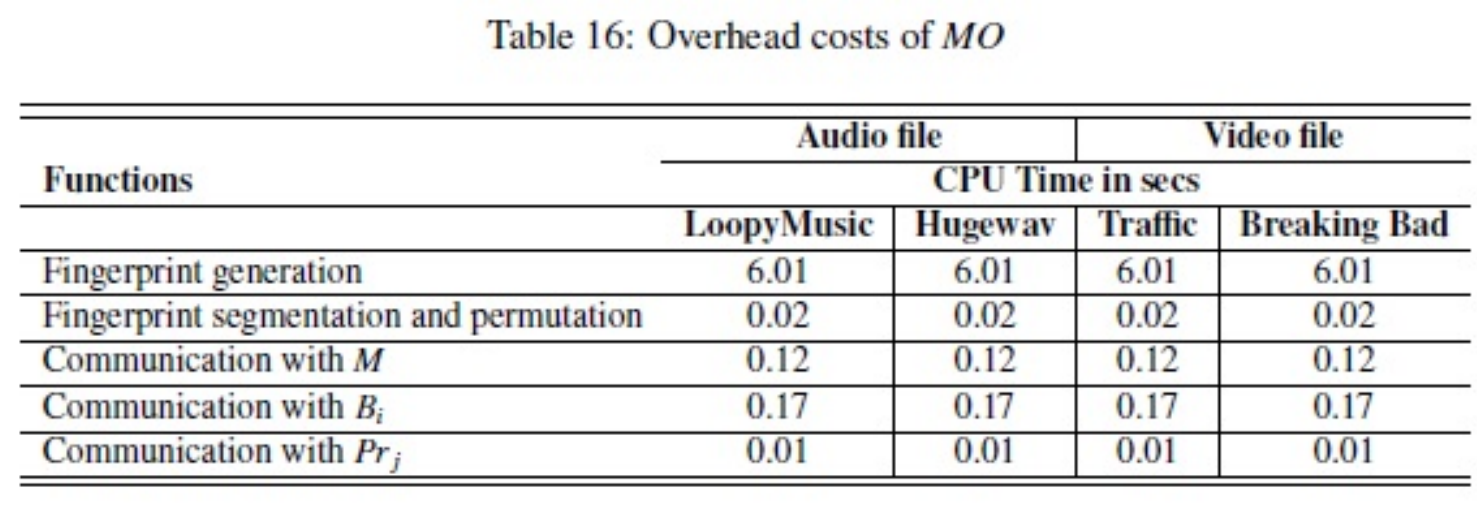}
	\label{fig:fig20}
\end{figure}

In Table 16, we present the overhead of \textit{MO} for ``LoopyMusic.wav", ``Hugewav.wav", ``Traffic.avi" and ``Breaking Bad.mp4". In the \textit{BF} distribution protocol, \textit{MO} is responsible for the generation of the collusion-resistant fingerprint $f_i$, the decryption of the received permutation keys from $B_i$, the segmentation and permutation of $f_i$, the transfer of encrypted session and permutation keys to $M$ and the allocation of the permuted fingerprint to $Pr_j$. It can be seen, from the table, that the time taken to generate $f_i$ is constant in each case. Similarly, the time taken by \textit{MO} to decrypt the permutation keys, segment and permute $f_i$, transfer the keys to $M$ and allocate the permuted $f_i$ to $Pr_j$ is constant in all cases. For a single file request, it takes \textit{MO} only $6.33$ seconds to perform all the desired functions. Thus, in case of multiple file requests, these tasks can be executed in parallel, therefore keeping the minimal overhead for \textit{MO}.

\begin{figure}[ht]
	\centering
	\includegraphics[width=14cm]{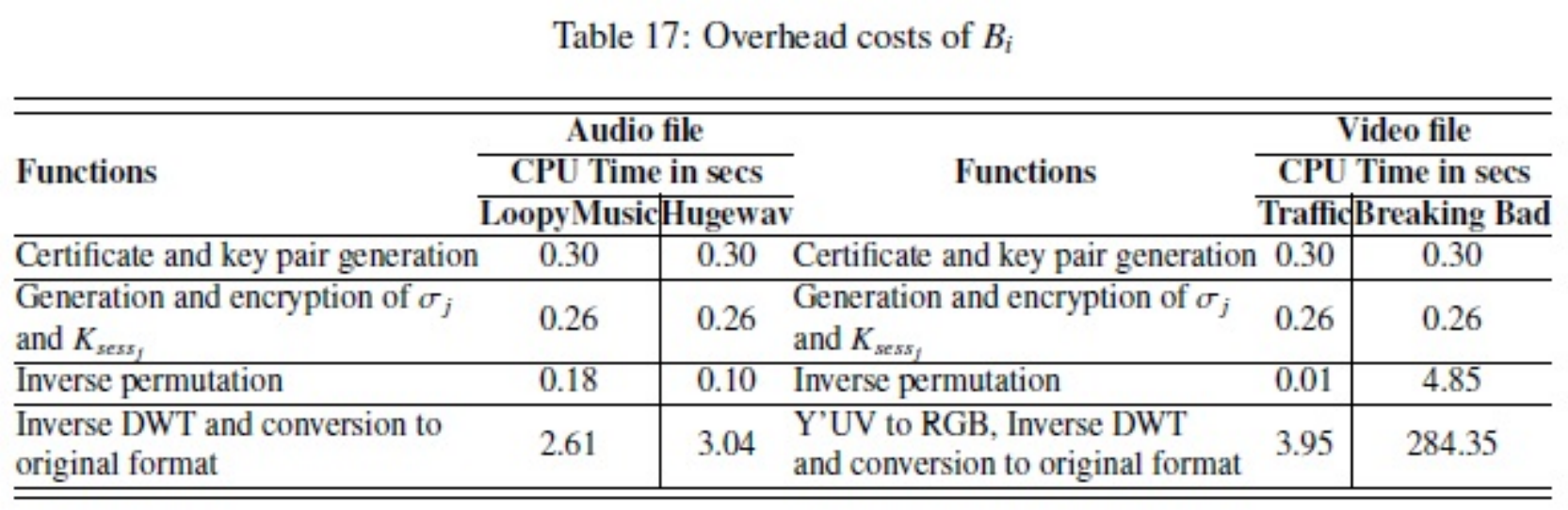}
	\label{fig:fig21}
\end{figure}

In the \textit{BF} distribution protocol, $B_i$ is responsible for the generation of an anonymous key pair, a certificate, permutation and session keys, and the file reconstruction at his/her end. Table 17 presents the overhead costs of $B_i$ for `` LoopyMusic.wav", `` Hugewav.wav", ``Traffic.avi" and ``Breaking bad.mp4". From Table 17, it can be seen that the time taken to generate the certificate, key pair, permutation and session keys, and the encryption of the keys, is constant in each case. Hence, for varying audio and video file sizes, $B_i$ can perform these operations in $0.56$ seconds. The inverse permutation and the decryption of the coefficients are also executed within a few seconds for both audio and video files. For an audio file, the inverse DWT and the conversion to original format does not require much time as compared to a video file. In a video file, the inverse DWT, Y'UV to RGB conversion, frame writing and conversion of frames to a video format takes a relatively longer time.

\begin{figure}[ht]
	\centering
	\includegraphics{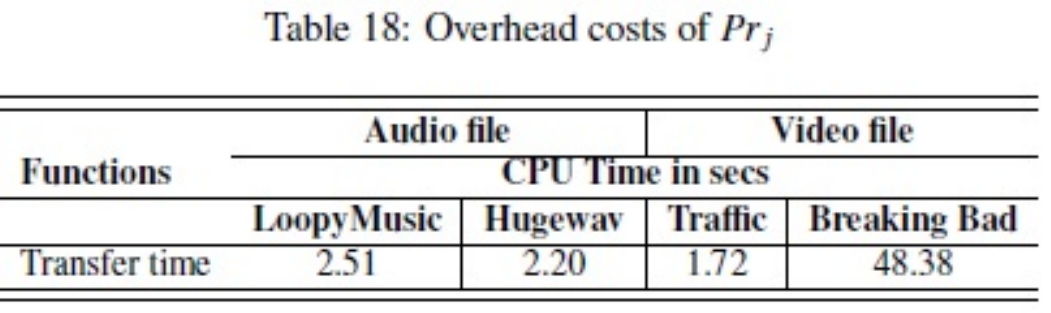}
	\label{fig:fig22}
\end{figure}

The proxy peers ($Pr_j$) are solely responsible for transferring the encrypted and permuted approximation coefficients from $M$ to $B_i$ in the \textit{BF} distribution protocol. Table 18 presents the overhead cost of $Pr_j$ for two audio and two video files. In case of a large-sized \textit{BF}, we can see that $Pr_j$ can deliver the contents in a secure manner between $M$ and $B_i$ in less than a minute. For small-sized files, the time taken by $Pr_j$ is approximately three seconds. Thus, it can be easily said that the overhead contributed by $Pr_j$ in the \textit{BF} distribution protocol is negligible.

\subsection{Response time}
\label{sec5.4}

Table 19 summarizes the response time for an audio file ``LoopyMusic.wav" and a video file ``Breaking bad.mp4". The response time is calculated as the time taken in \textit{BF} distribution from $M$ to $B_i$ through $Pr_j$ in the presence of \textit{MO}, the complete transfer of \textit{SF} from the providing peer to the requesting peer through an anonymous path, and the reconstruction of a file at $B_i$'s end. In Table 19, we also compare the response time of PSUM for ``LoopyMusic.wav" and ``Breaking bad.mp4" with the published response time results of \cite{qmr15} framework. The time taken in \textit{BF} generation and its distribution from $M$ to a buyer using an asymmetric fingerprinting protocol based on additive homomorphic cryptosystem, \textit{SF} distribution and file reconstruction is shown in Table 19. For \textit{BF} generation in the system of \cite{qmr15}, the fingerprint $f_i$ is embedded only once into the approximation coefficients $a_3$ of audio, whereas for a video file, the approximation coefficients $a_4$ of only two key frames are embedded with $f_i$. The remaining approximation coefficients are encrypted block-by-block using public-key cryptography. However, in the fingerprinting protocol of PSUM, all coefficients $a_3/a_4$ of an audio file and all coefficients $a_4$ of a video file are embedded with $f_i$. For \textit{SF} distribution, the execution time is constant in both systems. However, the file reconstruction process of PSUM is different from the framework of \cite{qmr15}, since, in the latter system, the buyer has to perform a public-key decryption of the content in comparison to PSUM, where the buyer only needs to perform an inverse permutation and a symmetric decryption of the content. It can be seen, from the last column of Table 19, that the total file (\textit{BF} and \textit{SF}) distribution time in PSUM is comparatively shorter than the distribution time using the scheme of \cite{qmr15}. The content distribution time in PSUM is better due to the use of symmetric encryption, instead of applying complex cryptographic protocols, such as homomorphic encryption, for \textit{BF} generation and distribution. Thus, it is evident, from Table 19 that the proposed fingerprinting protocol of PSUM permits to reduce the communication burden of secure fingerprinted content distribution.

\begin{figure}[ht]
	\centering
	\includegraphics[width=14cm]{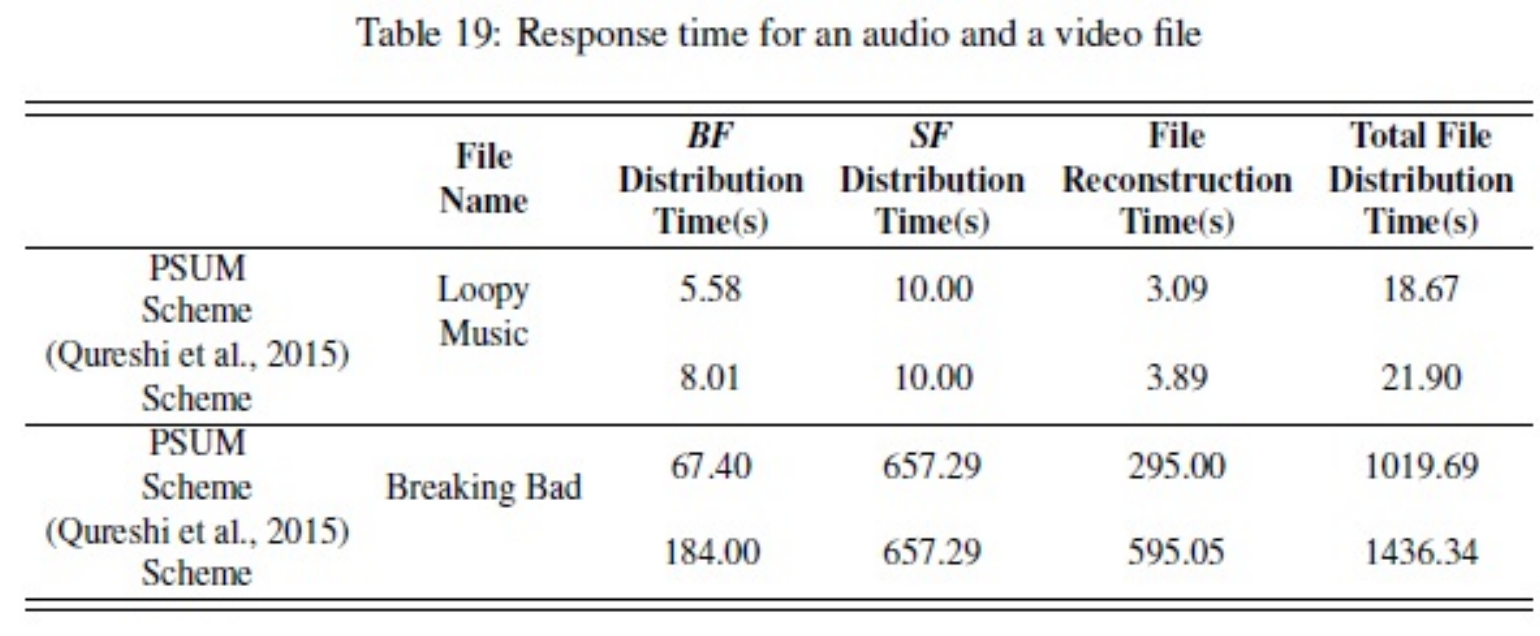}
	\label{fig:fig23}
\end{figure}

\subsection{Cryptographic costs}
\label{sec5.5}

Cryptographic algorithms are applied in PSUM to ensure the desired level of security, privacy and accountability. The cryptographic algorithms are implemented in C++ using the \cite{NTL} library. Table 20 shows the CPU execution time of each cryptographic block for achieving the desired security for the audio file ``LoopyMusic.wav". It is evident, from the table, that the anonymous paths construction and authentication through these paths is the most expensive cryptographic operation in PSUM. However, in achieving anonymity in P2P systems, there is always a cryptographic overhead. This overhead is due to encryption and decryption, insertion of fake traffic and increasing the routing path to provide anonymity between two communicating users. Still, the overhead of the authentication in PSUM is better due to the use of symmetric encryption, instead of applying asymmetric encryption. In PSUM, public-key cryptography is restricted to the generation of an anonymous certificate and a key pair, and encrypting the small-sized session and permutation keys during \textit{BF}   distribution from $M$ to $B_i$ through $Pr_j$ in the presence of \textit{MO}. Table 20 also shows the CPU execution time of each cryptographic block used in the system proposed by \cite{qmr15} for ``LoopyMusic.wav".

\begin{figure}[ht]
	\centering
	\includegraphics{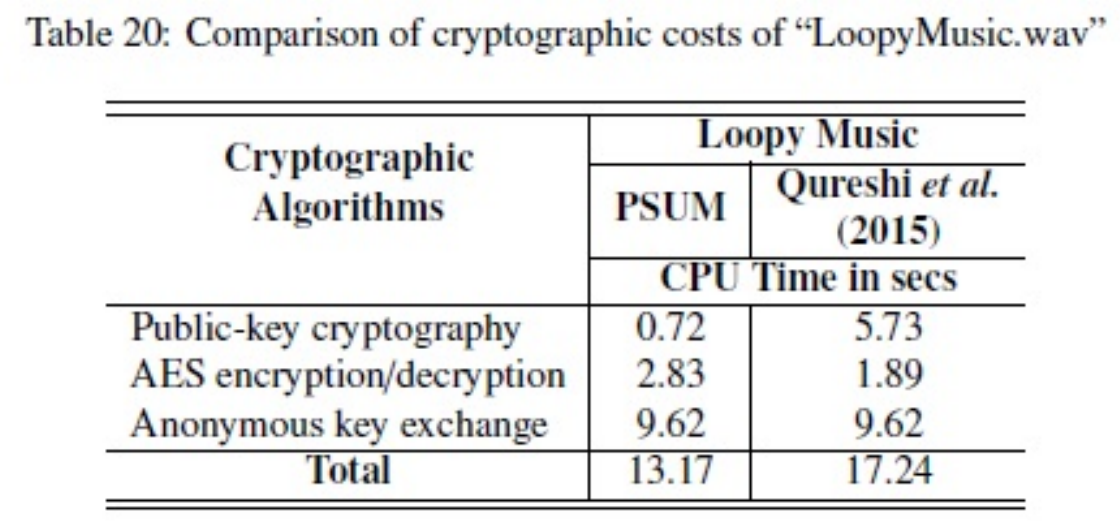}
	\label{fig:fig24}
\end{figure}

It is evident, from Table 20, that the cryptographic costs of the system by \cite{qmr15} are relatively larger than PSUM. The lower cryptographic overhead of PSUM is due to the use of the AES-128 symmetric-key algorithm to encrypt the pre-computed base files of multimedia files instead of using the $1024$-bit public-key encryption to produce the fingerprinted base file. The $1024$-bit Paillier encryption of a base file contributes to the high cryptographic costs of the system by \cite{qmr15}. The cryptographic overheads due to the anonymous AKE protocol are constant in the compared systems, since the number of tail nodes and onion paths between two communicating peers are considered fixed in both systems. Thus, the difference lies in how the cryptographic overhead is reduced in the base file distribution protocol. Thus, it can be said that PSUM provides a more efficient solution in terms of cryptographic costs.

\section{Conclusion}
\label{sec6}

We have proposed an efficient P2P content distribution system, i.e. a system that provides both copyright and privacy protection to the merchant and the buyers, respectively. In contrast to the known asymmetric fingerprinting schemes, which use homomorphic encryption to embed a fingerprint into a multimedia content and inflict high computational and communication burden on a merchant, our system lessens this cost for the merchant (and buyers) by only sending a small-sized base file composed of pre-computed fingerprinted information bits through proxies to the buyers. The main achievements of the PSUM are: (1) buyer security and privacy preservation, (2) collusion resistance and piracy tracing due to the use of \cite{Nu10} collusion-resistant fingerprinting codes, and (3) efficient content distribution by avoiding multi-party security protocols, bit commitments and public-key cryptography of the content. The security and performance analysis shows that PSUM is secure, privacy-preserving and efficient, compared to prior art.

\section*{Acknowledgment}                             
This work was partly funded by the Spanish Government through grants TIN2011-27076-C03-02 ``CO-PRIVACY'' and TIN2014-57364-C2-2-R ``SMARTGLACIS''.

\bibliographystyle{elsarticle-harv}
\bibliography{database}

\end{document}